\documentclass[a4paper,USenglish]{lipics-v2021}

\usepackage{amsmath}
\usepackage{amssymb}
\usepackage{booktabs} 
\usepackage{caption} 
\usepackage{subcaption} 
\usepackage{graphicx}
\usepackage{pgfplots}
\usepackage[all]{nowidow}
\usepackage[utf8]{inputenc}
\usepackage{tikz}
\usetikzlibrary{er,positioning,bayesnet}
\usepackage{multicol}
\usepackage[ruled,vlined]{algorithm2e}
\usepackage{hyperref}
\newcommand{\ignore}[1]{}

\usepackage{comment}

\usepackage{mathtools}

\DeclarePairedDelimiterX\Set[1]{\lbrace}{\rbrace}%
 {  #1 }

\definecolor{blue}{HTML}{1F77B4}
\definecolor{orange}{HTML}{FF7F0E}
\definecolor{green}{HTML}{2CA02C}

\pgfplotsset{compat=1.14}

\def\REVIEW{1}  

\if \REVIEW 1
    \newcommand{\liron}[1]{{\color{blue}[Liron: #1]}}
    \newcommand{\joao}[1]{{\color{red}[Joao: #1]}}
    \newcommand{\petr}[1]{{\color{green}[Petr: #1]}}
    \newcommand{\stefan}[1]{{\color{brown}[Stefan: #1]}}
\else
    \newcommand{\liron}[1]{}
    \newcommand{\joao}[1]{}
    \newcommand{\petr}[1]{}
    \newcommand{\stefan}[1]{}
\fi

    \newcommand{\hide}[1]{}

\newcommand{\hashdef}{\mathit{SLASH}}
\newcommand{\Sign}{\textit{sign}}
\newcommand{\Verify}{\textit{verify}}

\newcommand{\tx}{\mathit{tx}}
\newcommand{\seq}{\mathit{seq}}
\newcommand{\Id}{\mathit{id}}

\newcommand{\Iff}{\textbf{iff}}

\newcommand{\NOTIFY}{\textit{NOTIFY}}

\newcommand{\Min}{\mathit{min}}
\newcommand{\Max}{\mathit{max}}

\newcommand{\Hist}{\mathit{Hist}}

\newcommand{\REPLY}{\textit{REPLY}}

\newcommand{\Distance}{\textit{distance}}

\newcommand{\broadcast}{\mathit{broadcast}}
\newcommand{\deliver}{\mathit{deliver}}
\newcommand{\Validity}{\mathit{Validity}}
\newcommand{\Consistency}{\mathit{Consistency}}
\newcommand{\Totality}{\mathit{Totality}}
\newcommand{\Integrity}{\mathit{Integrity}}
\newcommand{\rS}{\textbf{$[S]$}}
\newcommand{\rPi}{\textbf{$[\Pi]$}}
\newcommand{\rW}{\textbf{$[W]$}}

\newcommand{\LPRB}{\textbf{L-PRB}}

\newcommand{\ECHO}{\mathit{ECHO}}
\newcommand{\READY}{\mathit{READY}}
\newcommand{\VALIDATE}{\mathit{VALIDATE}}
\newcommand{\WBB}{\textbf{WBB}}
\newcommand{\GetAllWitnesses}{\mathit{getPotWitnesses}}
\newcommand{\GetOwnWitnesses}{\mathit{getOwnWitnesses}}
\newcommand{\SetTimeout}{\mathit{setTimeout}}
\newcommand{\Timeout}{\mathit{timeout}}
\newcommand{\RECOVER}{\mathit{RECOVER}}
\newcommand{\RecHist}{\mathit{recHist}}

\newcommand{\Binomial}{\textit{Binomial}}

\newcommand{\SH}{\textbf{SH}}
\newcommand{\Split}{\textit{Split}}
\newcommand{\Recover}{\textit{Recover}}

\newcommand{\Checkpoint}{\textit{checkpoint}}
\newcommand{\Secret}{\textit{secret}}

\newcommand{\Convicted}{\textit{convicted}}

\newcommand{\REVEAL}{\mathit{REVEAL}}
\newcommand{\DONE}{\mathit{DONE}}
\newcommand{\FAILED}{\mathit{FAILED}}

\newcommand{\LOneDist}{\textit{1-Dist}}
\newcommand{\Border}{\textit{border}}
\newcommand{\AVG}{\mathit{Avg}}
\newcommand{\MixTime}{\textit{MixingTime}}
\newcommand{\TotalVar}{\textit{TotalVariationDistance}}

\newcommand{\DefineWitness}{\mathit{defineWitnesses}}

\newcommand{\powset}[1]{\mathbb{P}\left(#1\right)}

\SetKwComment{Comment}{/* }{ */}
\SetKwProg{Guard}{upon}{:}{}
\SetKwProg{Operation}{operation}{:}{}
\SetKwProg{Elseif}{else if }{ then }{}

\newcommand{\myparagraph}[1]{\vspace{3.5pt}\noindent \textbf{#1}}

\author{Veronika Anikina}{ITMO University}{iamveronikanikina@gmail.com}{}{}

\author{João Paulo Bezerra}{LTCI, Télécom Paris, Institut Polytechnique de Paris}{joaopaulo.bezerra@telecom-paris.fr}{}{}

\author{Petr Kuznetsov}{LTCI, Télécom Paris, Institut Polytechnique de Paris}{petr.kuznetsov@telecom-paris.fr}{}{}

\author{Liron Schiff}{Akamai}{schiff.liron@gmail.com}{}{}

\author{Stefan Schmid}{Technische Universität Berlin}{schmiste@gmail.com}{}{}

\Copyright{V. Anikina, J. P. Bezerra, P. Kuznetsov, L. Schiff, S. Schmid}

\authorrunning{V. Anikina, J. P. Bezerra, P. Kuznetsov, L. Schiff, S. Schmid}

\nolinenumbers

\begin{document}


\title{Dynamic Probabilistic Reliable Broadcast}
%
%
%
\maketitle              
\begin{abstract}
Byzantine reliable broadcast is a fundamental primitive in distributed systems that allows a set of processes to agree on a message broadcast by a dedicated process, even when some of them are malicious (Byzantine).
It guarantees that no two correct processes deliver different messages, and if a message is delivered by a correct process, every correct process eventually delivers one.
Byzantine reliable broadcast protocols are known to scale poorly, as they require $\Omega(n^2)$ message exchanges, where $n$ is the number of system members.
The quadratic cost can be explained by the inherent need for every process to relay a message to every other process.

In this paper, we explore ways to overcome this limitation, by casting the problem to the probabilistic setting.
We propose a solution in which every broadcast message is validated by a small set of \emph{witnesses}, which allows us to maintain low latency and small communication complexity.
In order to tolerate the \emph{slow adaptive adversary}, we dynamically select the witnesses through a novel \emph{stream-local} hash function: given a stream of inputs, it generates a stream of output hashed values that adapts to small deviations of the inputs. 
Our performance analysis shows that the proposed solution exhibits significant scalability gains over state-of-the-art protocols.
\end{abstract}

\section{Introduction}
Modern distributed computing systems are expected to run in extremely harsh conditions.
Besides communicating over weakly synchronous, or even purely asynchronous communication networks, 
the processes performing distributed computations may be subject to failures: from hardware crashes to security attacks or malicious (Byzantine) behavior. 
In these environments, ensuring that a system never produces \emph{wrong} outputs (\emph{safety} properties) and, at the same time, makes progress by producing \emph{some} outputs is extremely challenging.
The distributed computing literature reveals a plethora of negative results, from theoretical lower bounds and impossibility results to empirical studies, that exhibit fundamental scalability limitations.  

Efficient protocols that tolerate Byzantine failures are in high demand.
Let us consider \emph{cryptocurrencies}, by far the most popular decentralized application nowadays. 
Originally, cryptocurrency systems were designed on top of consensus-based blockchain protocols~\cite{bitcoin,ethereum}. 
But consensus is a notoriously hard synchronization 
problem~\cite{flp,dls,pbft,CHT96}. 
It came as a good news that we do not need consensus to implement a cryptocurrency~\cite{at2-cons,Gup16}, which
gave rise to asynchronous, \emph{consensus-free} cryptocurrencies~\cite{astro-dsn,fastpay,pastro21disc} that confirmed to exhibit significant performance gains over the consensus-based protocols.
At a high level, these implementations replace consensus with (Byzantine) \emph{reliable broadcast}~\cite{bra87asynchronous}, where a designated sender broadcasts a message so that no two correct processes deliver different messages (\emph{consistency}), either all correct processes deliver a message or none does (\emph{totality}), and if the sender is correct, all correct processes eventually deliver the broadcast message (\emph{validity}).

Starting from the classical Bracha's algorithm~\cite{br85acb}, Byzantine reliable broadcast algorithms~\cite{ma97secure,MR97srm,toueg-secure} are known to scale poorly, as they typically have $O(n)$ per-process communication complexity, where $n$ is the number of processes.
This can be explained by their use of  \emph{quorums}~\cite{ma97bqs,quorum-systems}, i.e., sets of  processes that are large enough (typically more than $2/3n$) to ensure that any two such sets have at least one correct process in common.
By relaxing the quorum-intersection requirement to only hold with high probability, the per-process communication complexity can be reduced to $O(\sqrt{n})$~\cite{probabilisticquorums}.    
Guerraoui et al.~\cite{scalable-rb-disc} describe a probabilistic broadcast protocol that replaces quorums with randomly selected \emph{samples}. 
This gossip-based broadcast consists of three phases, where in each phase involves communication with  with a small (of the order $O(\log{n})$), randomly  selected set of processes (a sample), which would give $O(\log{n})$ per-process communication cost.    
It can be shown that assuming a static adversary and an underlying uniform random sampling mechanism, the protocol can be tuned to guarantee almost negligible probabilities of failing the properties of reliable broadcast. %

In this paper, we take a step forward, by introducing a probabilistic reliable-broadcast protocol that tolerates an adaptive adversary, incurs even lower communication overhead.
Our protocol replaces samples with \emph{witness sets}: every broadcast message is assigned with a dynamically selected small subset of processes that we call the \emph{witnesses} of this message.

The witnesses are approached by the receivers to check if no other message has been issued by the same source and with the same sequence number. 
The processes select the witness set by applying a novel \emph{stream-local} hash function to the current \emph{random sample}.
The random sample is a set of random numbers that the system participants periodically generate and propagate throughout the system, and we ensure that "close" random samples induce "close" witness sets.   
To counter a dynamic adversary manipulating the random samples and, thus, the witness sets in its favor, 
the random numbers are \emph{committed} in advance using a secret-sharing mechanism~\cite{reed1960polynomial,das2021asynchronous}.    
The committed random numbers are \emph{revealed} only after a certain number of broadcast instances, which is a parameter of the security properties of our protocol.      
As a result, the protocol is resistant against a \emph{slowly adaptive} adversary.
We model the evolution of the random samples as an ergodic Markov process: the time for the adversary to corrupt a process is assumed to be much longer  than the \emph{mixing time} of a carefully defined random walk in a multi-dimensional space.     
Intuitively, even if the adversary introduces a biased value instead of a random one, by the time the value is used, the distribution of the random sample is close to uniform and the benefits of the bias are lost.

We argue that the desired safety properties of Byzantine reliable broadcast can be achieved under small, $O(\log n)$, witness sets.
When the communication is close to synchronous, which we take as a common case in our performance analysis, the divergence between the random samples evaluated by different processes and, thus, the witness sets for the given broadcast event, is likely to be very small.
Thus, in the common case, our broadcast protocol maintains $O(n\log n)$ communication complexity, or $O(\log n)$ per node, similar to sample-based gossiping~\cite{scalable-rb-disc}, by additionally exhibiting \emph{constant} latency.  

The current ``amount of synchrony'' may negatively affect the liveness properties of our algorithm: the less synchronous is the network,
the less accurate may the witness set evaluation become and, thus, the longer is the delivery of a broadcast message.
Notice that we do not need the processes to perfectly agree on the witnesses for any particular event: thanks to the use of stream-local hashing, a \emph{sufficient} overlap is enough.
To compensate for liveness degradation when the network synchrony weakens, i.e., the variance of effective message delays gets higher, we propose a \emph{recovery} mechanism relying on the classical (quorum-based) reliable broadcast~\cite{bra87asynchronous}.  

Our comparative performance analysis shows that throughput and latency of our protocol scales better than earlier protocols~\cite{bra87asynchronous,scalable-rb-disc}, which makes its use potentially attractive in large-scale decentralized services~\cite{astro-dsn,fastpay}.           
In particular, we pinpoint two potential applications of our broadcast protocol: an efficient, broadcast-based  cryptocurrency  and a generic lightweight accountability mechanism~\cite{peerreview}. 

The rest of the paper is organized as follows. 
In Section~\ref{sec:model}, we describe the system model and in 
Section~\ref{sec:probBroadcast}, we formulate the problem of probabilistic reliable broadcast and present our baseline witness-based protocol. 
In Section~\ref{sec:witnessOracle}, we extend the baseline protocol to implement probabilistic broadcast.
In Section~\ref{sec:security}, we analyze the security properties of our protocol and  
we present the outcomes of our performance analysis. 
In Section~\ref{sec:optimistic}, we sketch a recovery mechanism that can be used to complement our baseline witness-based protocol. 
In Section~\ref{sec:relwork} we discuss related work and conclude the paper in Section~\ref{sec:conclusion}. 
Proofs and technical details of the secret-sharing and recovery schemes, as well as the overview of potential applications of our protocol are delegated to the appendix. 

\section{System Model}
\label{sec:model}

\myparagraph{Processes.}
A system is composed of a set $\Pi$ of \emph{processes}.
Every process is assigned an \emph{algorithm} (we also say \emph{protocol}).
Up to $f < |\Pi|/3$ processes can be corrupted by the adversary.
Corrupted processes might deviate arbitrarily from the assigned algorithm, in particular they might  prematurely stop sending messages.
A corrupted process is also called \emph{faulty} (or \emph{Byzantine}), otherwise we call it \emph{correct}.

We assume a slow adaptive adversary: it decides which processes to corrupt depending on the execution, but it takes certain time for the corruption to take effect.
When selecting a new process $p$ to corrupt at a given moment in the execution, the adversary can have access to $p$'s private information and control its steps only after every other correct process has terminated $\Delta$ protocol instances, where $\Delta$ is a predefined parameter.\footnote{In this paper we consider broadcast protocols (Section~\ref{sec:probBroadcast}), so that an instance terminates for a process when it delivers a message.}
In addition, we assume that previously sent messages by $p$ cannot be altered or suppressed.

\myparagraph{Channels.}
Every pair of processes communicate through authenticated reliable channels: messages are signed and the channel does not create, drop or duplicate messages.
We assume that the time required to convey a message from one correct process to another is much smaller than the time required for any individual correct process to terminate in $\Delta$ instances.\footnote{The time to process $\Delta$ instances can be in range of hours or days (see Section~\ref{sec:security}).}

\myparagraph{Cryptography.} We use hash functions and asymmetric cryptographic tools: a pair public-key/private-key is associated with every process in $\Pi$~\cite{cachin2011introduction}.  
The private key is only known to its owner and can be used to produce a \emph{signature} for a statement, %
while the public key is known by all processes and is used to \emph{verify} that a signature is valid. 
We assume a computationally bounded adversary: no process can forge the signature for a statement of a benign process.
In addition, signatures satisfy the uniqueness property: for each public-key $pk$ and a message $m$, there is only one valid signature for $m$ relative to $pk$.
The hash functions are modeled as a \emph{random oracle}.

\section{Probabilistic Reliable Broadcast}
\label{sec:probBroadcast}

The \emph{Byzantine reliable broadcast} abstraction~\cite{bra87asynchronous,cachin2011introduction} exports operation $\broadcast(m)$, where $m$ belongs to a message set $\mathcal{M}$, and  produces callback $\deliver(m')$, $m'\in\mathcal{M}$.
Each instance of reliable broadcast has a dedicated source, i.e., a single process broadcasting a message.
In any execution with a set $F$ of Byzantine processes,
the abstraction guarantees the following properties: 

\begin{itemize}
    \item $(\Validity)$ If the source is correct and invokes $\broadcast(m)$, then every correct process eventually delivers $m$.
    \item $(\Consistency)$ If $p$ and $q$ are correct and deliver $m$ and $m'$ respectively, then $m = m'$.
    \item $(\Totality)$ If a correct process delivers a message, then eventually every correct process delivers a message.
    \item $(\Integrity)$ If the source $p$ is correct and a correct process delivers $m$, then $p$ previously broadcast $m$.
\end{itemize}

\myparagraph{Long-lived reliable broadcast.} In a \emph{long-lived} execution of reliable broadcast, each process maintains a history of delivered messages and can invoke $\broadcast$ an unbounded number of times,
but correct processes behave sequentially, i.e., wait for the output of a $\broadcast$ invocation before starting a new one.
The abstraction can easily be implemented using an instance of reliable broadcast for each message, by attaching to it the source's $\Id$ and a sequence number~\cite{bra87asynchronous}.

\myparagraph{Probabilistic broadcast.}
Instances of reliable broadcast can also be probabilistic~\cite{scalable-rb-disc}, in which case there is a probability for each instance that the protocol does not satisfy some property (e.g. violates $\Consistency$).
If one uses instances of probabilistic reliable broadcast for building a long-lived abstraction, the probability of failure converges to 1 in an infinite run (assuming processes broadcast messages an infinite number of times).
We therefore consider the \emph{expected failure time} of a \emph{long-lived probabilistic reliable broadcast} ($\LPRB$) as the expected number of broadcast instances by which the protocol fails.

\begin{definition}[Long-lived Probabilistic Reliable Broadcast]
An $\epsilon$-Secure $\LPRB$ has an expected time of failure (average number of instances until some property is violated) of $1/\epsilon$ instances.
\end{definition}

\subsection{Protocol Description}
\myparagraph{Witness-Based Protocol.}
We first present an algorithm that implements Byzantine reliable broadcast using a distributed \emph{witness oracle} $\omega$.
Intuitively, every process $p_i$ can query its local oracle module $\omega_i$ to map each \emph{event} $e=(id,seq)$ (a pair of a process identifier and a sequence number) to a set of processes that should \emph{validate} the $seq$-th event of process $id$.
The oracle module $\omega_i$ exports two operations: $\GetAllWitnesses(\Id,\seq)$, which returns a set $V_i$ of processes potentially acting as witnesses for the pair $(\Id,\seq)$, and $\GetOwnWitnesses(\Id,\seq)$, that returns a set $W_i \subseteq V_i$ of witnesses particular to $p_i$, referred to as $p_i$'s \emph{witness set}.

We now describe an algorithm that uses $w$ to implement Byzantine reliable broadcast,
a variation of Bracha's algorithm~\cite{bra87asynchronous} that instead of making every process gather messages from a\emph{quorum}, we delegate this task to the witnesses.\footnote{A quorum is a subset of processes that can act on behalf of the system. A Byzantine quorum~\cite{ma97bqs} is composed of $q = \lfloor \frac{n+f}{2} \rfloor + 1$ processes, for a system with $n$ processes in which $f$ are Byzantine.}
Each process waits for replies from a threshold $k$ of its witnesses in $W_i$ to advance to the next protocol phase.
However witness sets can differ, so messages are sent to $V_i$ to guarantee that every process acting as a witness can gather enough messages from the network.
The protocol maintains correctness as long as for each correct process $p_i$, there are at least $k$ correct witnesses and at most $k-1$ faulty witnesses in $W_i$. Moreover, $V_i$ should include the witness set of every other correct process.
In Section~\ref{sec:witnessOracle}, we describe a method to select $k$ and a construction of $\omega$ that satisfies these with very large expected failure time.

The pseudo-code for a single instance of Byzantine reliable broadcast (parameterized with a pair $(\Id,\seq)$) is presented in Algorithm~\ref{alg:wbb}.
Here we assume the set of participants $\Pi$ ($|\Pi| = n$) to be static: 
the set of processes remains the same throughout the execution.
$f$ denotes the number of faulty nodes tolerated in $\Pi$ (see~\ref{subsec:correctness}).

\begin{algorithm}
\caption{\textbf{W}itness \textbf{B}ased \textbf{B}roadcast}
\label{alg:wbb}
\BlankLine

\nl$\rPi$ \Guard{init - $(\Id,\seq)$}{
    \nl $V_i \gets \omega_i.\GetAllWitnesses(\Id,\seq)$\;
    \nl $W_i \gets \omega_i.\GetOwnWitnesses(\Id,\seq)$\;
}
\BlankLine

$\nl\rS$ \Operation{$\broadcast(m)$}{
    \nl send $\langle \NOTIFY, m, S \rangle$ to every $p \in \Pi$\; 
} 
\BlankLine

$\nl\rPi$ \Guard{receiving $\langle \NOTIFY, m, S \rangle$}{ 
    \nl send $\langle \ECHO, m, \Pi \rangle$ to every $w \in V_i$\; 
}
\BlankLine

$\nl\rW$ \Guard{receiving $\lfloor \frac{n+f}{2} \rfloor + 1$ $\langle \ECHO, m, \Pi \rangle$ or $f+1$ $\langle \READY, m, \Pi \rangle$ messages}{
    \nl send $\langle \READY, m, W \rangle$ to every $p \in \Pi$\; \label{line:readyTrigger}
}
\BlankLine

$\nl\rPi$ \label{line:phase1k} \Guard{receiving $\langle \READY, m, W \rangle$ from $k$ witnesses in $W_i$} {
    \nl send $\langle \READY, m, \Pi \rangle$ to every $w \in V_i$\;
}
\BlankLine

$\nl\rW$ \Guard{receiving $\langle \READY, m, \Pi \rangle$ from $\lfloor \frac{n+f}{2} \rfloor + 1$ processes} {
    \nl send $\langle \VALIDATE, m \rangle$ to every $p \in \Pi$\;
}
\BlankLine

$\nl\rPi$ \label{line:phase2k} \Guard{Upon receiving $\langle \VALIDATE, m \rangle$ from $k$ witnesses in $W_i$} {
    \nl $\deliver(m,\Id,\seq)$\;  \label{line:wbbDeliver}
}
\BlankLine
\end{algorithm}

\myparagraph{Initialization.} At the source $s$, a single instance of \emph{Witness-Based Broadcast} ($\WBB$) parameterized with $(s,\seq)$ is initialized when $s$ invokes $\broadcast(m)$,
where $(s,\seq)$ is attached to $m$.
On the remaining processes, the initialization happens when first receiving a protocol message associated to $(s,\seq)$.
Upon initialization, processes sample $V_i$ and $W_i$ from $\omega_i$, which are fixed for the rest of the instance.

\myparagraph{Roles.} Each action is tagged with $\rS$, $\rW$ or $\rPi$, where $\rS$ is an action performed by the source, $\rW$ is performed by a process acting as witness and $\rPi$ by every process.
A process $p_i$ can take multiple roles in the same instance and always takes actions tagged with $[\Pi]$, but performs each action \emph{only once per instance}.
Moreover, $p_i$ (correct) acts as witness \textit{iff} $p_i \in V_i$.
We assume every broadcast $m$ to include the source's signature so that every process can verify its authenticity.

Algorithm~\ref{alg:longBroad} uses $\WBB$ as a building block to validate and deliver messages.
Clearly, if the validation procedure satisfies the reliable broadcast properties, then Algorithm~\ref{alg:longBroad} implements long-lived reliable broadcast. 

\begin{algorithm}
\caption{Long-Lived Reliable Broadcast}
\label{alg:longBroad}
\BlankLine

\textbf{Local Variables:} \\
$\seq \gets 0$\Comment*[r]{Sequence number of $p_i$}
$\Hist \gets \emptyset$\Comment*[r]{Delivered messages}
\BlankLine

\nl \Operation{$\broadcast(v)$}{
    \nl $\seq++$\;
    \nl $m \gets (v,p_i,\seq)$\;
    \nl $\WBB.\broadcast(m)$\;
}
\BlankLine

\nl \Guard{$\WBB.\deliver(m)$}{
    \nl $\Hist \gets \Hist \cup \{m\}$\;
}
\BlankLine

\end{algorithm}

\subsection{Protocol Correctness}
\label{subsec:correctness}
\myparagraph{Assumptions.} Consider an instance of $\WBB$ where $f < |\Pi|/3$ and for every $p_i$ correct:

\begin{enumerate}
    \item $W_i$ has at least $k$ correct witnesses and at most $k-1$ faulty witnesses; \label{enum:w}
    \item for every $p_j$ correct, $W_j \subseteq V_i$.\label{enum:v}
\end{enumerate}

Then the following theorem holds:

\begin{theorem}
\label{th:sketchCorrectness}
Algorithm~\ref{alg:wbb} implements Byzantine reliable broadcast.
\end{theorem}
\begin{proof} $\Validity$: Let $p_i$ be a correct process and assume that a correct source is broadcasting $m$.
Since for every $p_j$ correct $W_i \subseteq V_j$ (assumption \textbf{2}), all correct witnesses in $W_i$ receive $\lfloor \frac{n+f}{2} \rfloor + 1$ echoes and reply with $\READY$.
From assumption \textbf{1}, $W_i$ has at least $k$ correct witnesses that reply to $p_i$, which in turn sends its own $\READY$ back.
The validation phase follows similarly and $p_i$ delivers $m$ after receiving $k$ $\VALIDATE$ messages from witnesses.

$\Integrity$: $m$ is signed by the source and processes verify its authenticity, so a message broadcast is only delivered if the authentication is successful.
By assumption, the adversary cannot crack the private key of a correct process and forge signatures.

$\Consistency$:
Because $W_i$ has at most $k-1$ faulty processes, $p_i$ is guaranteed to receive a message from at least one correct process in lines~\ref{line:phase1k} and \ref{line:phase2k} before proceeding to a new phase.
A correct witness has to receive $\lfloor \frac{n+f}{2} \rfloor + 1$ $\langle \READY, m, \Pi \rangle$ in order to send $\langle \VALIDATE, m \rangle$.
Since $f < |\Pi|/3$, every pair of subsets of $\lfloor \frac{n+f}{2} \rfloor + 1$ processes intersects in at least one correct process, thus two correct witnesses cannot send $\VALIDATE$ for distinct messages (this would require a correct process to send $\READY$ for distinct messages).

If a correct process delivers $m$, than at least $\lfloor \frac{n+f}{2} \rfloor + 1$ processes sent $\langle \READY, m, \Pi \rangle$ to a correct witness, thus, at least $f+1$ correct processes sent $\langle \READY, m, \Pi \rangle$ to every witness.
Consequently from assumption \textbf{2}, correct witnesses receive $f+1$ readies for a message and are able to trigger line~\ref{line:readyTrigger} to send $\langle \READY, \cdot, W \rangle$.
In order to send $\READY$ without hearing from $f+1$ processes, witnesses gather echoes from $\lfloor \frac{n+f}{2} \rfloor + 1$ processes.
Similarly to the $\Consistency$ part of the proof, two correct witnesses cannot then send $\READY$ for distinct messages.
Finally, $\Totality$ holds from the fact that every $W_i$ has at least $k$ correct witnesses, which send $\langle \READY, m, \Pi \rangle$ to all processes.
\end{proof}

\myparagraph{Complexity.} For a single broadcast instance, the message complexity depends on the size of $V_i$ (the set of potential witnesses).
Let $|\Pi| = n$ and $v$ the expected size of $V_i$, the message complexity is $O(n \cdot v)$.
We assume that parameters for the witness-oracle (Section~\ref{sec:witnessOracle}) are chosen such that $v$ is $\Omega(\log n)$, resulting in a complexity of $O(n\log n)$. Moreover, $\WBB$ takes $5$ message delays to terminate with a correct source.

\section{Witness Oracle}
\label{sec:witnessOracle}

\begin{figure}[htbp]
  \caption{Illustration of how stream local hashing of similar histories ($S$ and $\hat{S}$) results in similar witness set selections ($W$ and $\hat{W}$).}
  \centerline{\includegraphics[width=9cm]{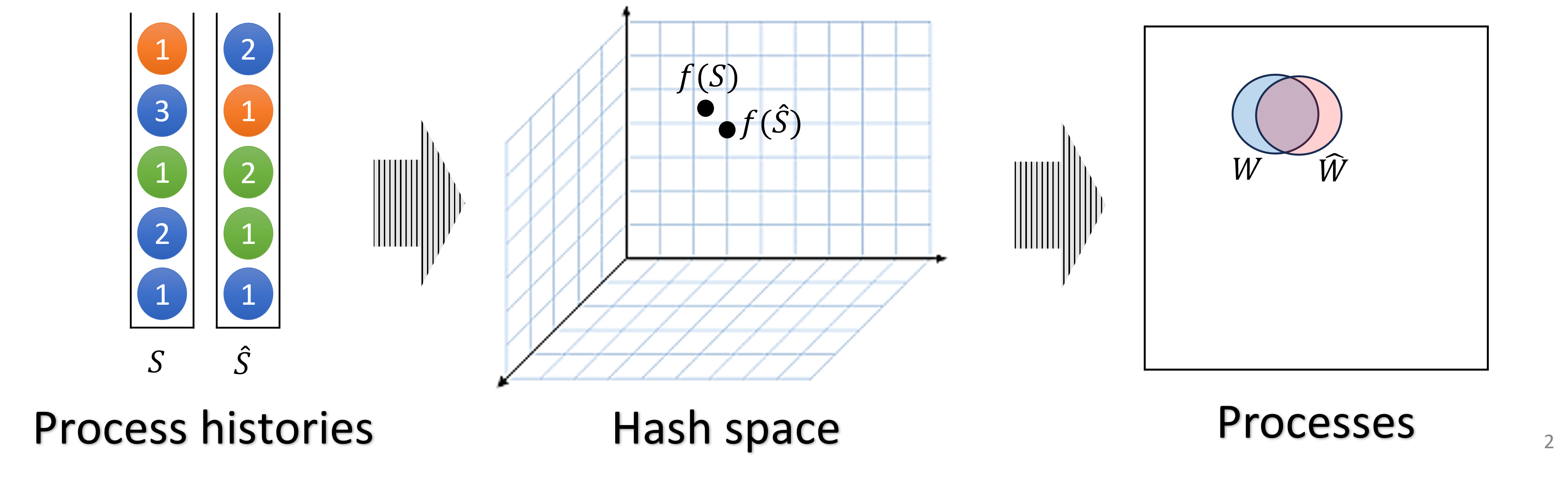}}
  \label{fig:illustration}
\end{figure}

The witness oracle is responsible for selecting a set of witnesses for each broadcast message. This selection should be unpredictable so that it is known only with close proximity to the time of receiving the broadcast message.
We implement this service locally on each node and require that for every broadcast message and for any pair of nodes, the locally selected witness sets will be similar.

In order to provide unpredictable outputs,
each node uses the hash of a history of random numbers that are jointly delivered with messages.
The witness set used in a particular broadcast instance is defined according to the current history, and remains the same (for that instance) after the first oracle call.

\myparagraph{Properties - Witness Oracle.} Let $W_i^l(\Id,\seq)$ and $V_i^l(\Id,\seq)$ be the outputs for $\omega_i.\GetOwnWitnesses(\Id,\seq)$ and $\omega_i.\GetAllWitnesses(\Id,\seq)$ respectively at the moment $p_i$'s history has $l$ elements.
The witness oracle construction satisfies:

\begin{itemize}
    \item (Witness Inclusion) For any pair of correct nodes $p_i$ and $p_j$, and instance $(\Id, \seq)$:
    $$
    W_i(\Id,\seq) \subseteq V_j(\Id,\seq)
    $$
    \item (Unpredictability) Let $S$ be the history of correct process $p$ and $w$ the average witness set size. Then for a given $\epsilon > 0$, there exists $L \in \mathbb{N}$ such that, for any process $q$ and $L' \geq L$:
    $$
    \Pr(q \in W^{l+L'}(\cdot,\cdot) \ | \ |S| = l) \leq \frac{1}{w} + \epsilon
    $$
\end{itemize}

\emph{Witness Inclusion} ensures that correct processes use close witness sets in the same instance. \emph{Unpredictability} hinders processes from accurately predicting witness sets: based on $p$'s history $S$ at any point in the execution, it is impossible to precisely determine whether any process $q$ will be selected as a witness after $L$ instances.

To support a high rate of broadcast messages, nodes do not wait or try to establish the same histories before hashing.
Instead, we use a novel \emph{locality sensitive} hashing scheme to ensure that small differences in histories will result in small differences in the selected witness set, as illustrated in Figure~\ref{fig:illustration}.

The history of random numbers can be regarded as a set. However, existing locality sensitive hashing algorithms for sets (such as those based on MinHash~\cite{broder1997resemblance}) are less sensitive to small differences with bigger sets,
which does not fit well the case of histories that are expected to grow indefinitely.
Note that hashing just a sliding window within histories 
may result with distinct nodes having different hash values even if their histories are the same, due to differences in the order that values can be added.

Another disadvantage of existing locality sensitive hashing algorithms is their vulnerability to messages crafted to manipulate the hash result. For example in MinHash based algorithms, adding an item with extremely small hash value will most likely keep the hash of the entire set constant for a long time regardless of insertions of new items. In our case, when using history hash for witness selection, it could allow the adversary to make the witness selection predictable.

\myparagraph{Properties - Hash Function.} In short, let $f: \powset{\{0,1\}^*} \to \mathbb{Z}_r^b$ be a stream local function that can hash a set of binary strings into a vector in $b$-dimensional $\mod r$ torus. We require $f$ to satisfy:

\begin{itemize}
    \item (Locality) Let $S_i,S_j \in \powset{\{0,1\}^*}$. There exists $\lambda > 0$ and $L \in \mathbb{N}$ such that:
    $$
    (|S_i \cup S_j - S_i \cap S_j| \leq L) \Longrightarrow \Distance(f(S_i),f(S_j)) \leq \lambda L
    $$
    For some pre-defined distance measure in $\mathbb{Z}_r^b$.
\end{itemize}

Next, we show a construction of a safe stream local hashing for histories ($\hashdef$), a hashing scheme that guarantees high degree of unpredictability and also similar outputs for similar histories. Later we show how $\hashdef$ can be used for witness selection.

\subsection{Constructing a safe stream local hashing of histories}
\label{subsec:localHash}
We consider a history as a set of binary strings and we assume the existence of a family of one-way functions $H_c=\left\{h_s:\{0,1\}^*\to\{0,1\}^c \right\}$. For example, for $c=256$, we can use $h_s(x) = SHA256(x\oplus s)$ where $s$ is a predefined random binary string with length at least $c$ and ``$\oplus$'' is the bit concatenation operator.

We define a family of $\hashdef$ functions $F_{r,b,H_c}=\left\{f_s:\powset{\{0,1\}^*}\to \mathbb{Z}_r^b \right\}$, where $b<2^c$. Each of these functions can hash a set of arbitrary binary strings into a vector in $b$-dimensional $\mod r$ torus and should be locality sensitive in the sense that sets with small differences should be hashed to vectors with small distance, where vector distance is defined as follows:

 \[ TorusDist_{r,b}(X, Y) := \Max ( \{ min(X_j-Y_j \mod{r}, Y_j-X_j \mod{r}) \ | \ j \in \{\ 0,\ldots,b-1 \} \} ) \]

Our construction of $\hashdef$ functions $F_{r,b,H_c}$ can be described as follows: each evaluation of $f \in F_{r,b,H_c}$ on a set $S\subseteq\{0,1\}^*$ defines a random walk in $\mathbb{Z}_r^b$ where each item $x$ in $S$ accounts for an independent random step based on the hash of $x$. More specifically:
$$
f(S) := \left<g\left(S_0\right), g\left(S_1\right), \ldots, g\left(S_{b-1}\right)\right>,
$$ where $S_y:=\{x\in S \ | \ \ h_s(x) \mod b = y\}$ and $g(V)=\sum_{v\in V} (-1)^{h_i(v) \div b} \mod r$.

The distance between any two sets is not affected by shared items while each non-shared item increases or decreases the distance by at most $1$.

\begin{theorem}[Locality]
\label{th:locality}
    Let $S,T \in \powset{\{0,1\}^*}$, then for any $f \in F_{r,b,H_c}$:
    \[TorusDist_{r,b}(f(S),f(T)) \leq |S \cup T - S \cap T|.\]
\end{theorem}

\begin{proof}
    Each element $x \in S$ can be mapped to a single $S_y$, in which it either increases or decreases $g(S_y)$ by $1$.
    Suppose that $S \cup T - S \cap T = \{x\}$ (so either $S = T \cup \{x\}$ or $T = S \cup \{x\}$).
    Let $h_s(x) \mod b = y$, then $f(S)$ is identical to $f(T)$ in all positions $0,\ldots,b-1$ except for $y$, where $f(S)[y] = f(T)[y] \pm 1$.
    So: \[TorusDist_{r,b}(f(S),f(T)) = 1 = |S \cup T - S \cap T|\]

    Now assume that $TorusDist_{r,b}(f(U),f(V)) \leq |U \cup V - U \cap V|$ for $|U \cup V - U \cap V| = L$, and let $S \cup T - S \cap T = \{x_1,\ldots,x_{L+1}\}$.
    Assume without loss of generality that $x_{L+1} \in S$, and let $S' = S - \{x_{L+1}\}$.
    Then by assumption, for any position $m$ in $0,\ldots,b-1$:
    \[min(f(S')[m]-f(T)[m] \mod{r}, f(T)[m]-f(S')[m] \mod{r}) \leq L\]

    Now let $h_s(x_{L+1}) \mod b = y$, then the inequalities above are satisfied for $f(S)$ and $f(T)$ at any position $0,\ldots,b-1$ except maybe for $y$, where:
    \[min(f(S)[y]-f(T)[y] \mod{r}, f(T)[y]-f(S)[y] \mod{r}) \leq L+1\]
    Therefore: $TorusDist_{r,b}(f(S),f(T)) \leq |S \cup T - S \cap T|$.
\end{proof}

Theorem~\ref{th:locality} implies that $\hashdef$ satisfies \emph{Locality} with $\lambda = 1$.

\myparagraph{Using \textit{SLASH} functions for witness set selection.}
We consider node ids to be binary strings of length $c$.
During the execution, each node $i$ maintains a view regarding the set of active node ids $\Pi_i \subseteq \{0,1\}^d$ and the history of delivered random numbers $R_i$.
In addition, a $\hashdef$ function $f_j$ is maintained for each node in $\Pi_i$.
All nodes are initialized with the same functions $f_j$ as well as the one-way function $h_j \in H_c$, so that for each particular $f_j$ the id of node $j$ is used as the seed for computing $h_j(x)$,
in order to make each step computed for different $f_j$ independent.

To select witness sets, nodes are also initialized with a distance parameter $d\in\mathbb{R}^+$.
To determine the set of witnesses $W_i(\Id,\seq)$, node $i$ computes $y_j = f_j(R_i)$ and then selects all nodes $j$ for which $y_j$ is at distance at most $d$ from the origin.
\begin{equation}\label{eq:witness-critertion}
    W_i= \{j \in \Pi_i \ | \ TorusDist_{r,b}(y_j,[0]) \leq d \}.
\end{equation}

Keeping a $\hashdef$ instance for each node does not add significant storage overhead, since we expect its size to be smaller than $c$ in number of bits (see Section~\ref{subsec:security}).
Moreover, the computational cost of updating $\hashdef$ should also be significantly smaller than verifying digital signatures.

\subsection{Secret Sharing}
Unpredictability is achieved with a secret sharing protocol, in which a random number is secretly shared by the source at the start of every instance.
In the reveal phase, the number is added to a local history and it is used to compute a random step in $\hashdef$.

We capitalize on the steady distribution of a random walk
which is the uniform distribution in a torus\footnote{To guarantee uniform distribution, the diameter $r$ needs to be odd~\cite{berestycki2016mixing}, we can trivially achieve this by skipping the last point (so the diameter of each dimension becomes $r-1$).} (this means that as we increase the number of steps, the distribution of $\hashdef$ converges to uniform).
The number of steps necessary to make the distribution of the random walk close to uniform is called the \emph{mixing time}.
For each new random number that is shared by a node, we delay its addition to the history by $\delta = mixTime/th_c$ steps, where $th_c$ is the fraction of the throughput generated by correct sources.\footnote{One can adjust $th_c$ based on the broadcast rate of the $n-f$ nodes with smallest rate. Alternatively one can wait until the total number of additions to the history, committed by the  $n-f$ nodes with smallest rate, has reached $mixTime$.} This guarantees that the adversary cannot use it's current state to issue carefully chosen numbers and make the outcome of $\hashdef$ close to any specific value.

To further prevent the adversary from biasing the outcome of $\hashdef$, we require each number to be generated by a verifiable source of randomness.
This is achieved using the signature scheme and a hash function.
Consider a random string $s$ known to all nodes.
To generate the random number $x$ of instance $(p_i,\seq)$, $p_i$ signs $s \oplus (p_i,\seq)$ and assigns to $x$ the hash of the resulting signature.
The signature for $s \oplus (p_i,\seq)$ is then used as proof that $x$ was correctly generated.
The random string $s$ used for $(p_i,\seq)$ is the random number $x'$ generated in the previous instance, and the original seed can be any agreed upon number.

We use Shamir's secret sharing scheme~\cite{shamir1979share}, the protocol's integration with $\WBB$ is described in Appendix~\ref{app:secret}.

\subsection{Witness Oracle - Correctness}
\label{subsec:oracleCorrect}
The detailed analysis for the oracle correctness can be found in Appendix~\ref{app:oracleCorrect}. To verify the conditions under which \textit{Witness Inclusion} and \textit{Unpredictability} are satisfied, we assume that:

\begin{itemize}
    \item The interval between the first time a correct process recovers a secret $x'$, and the last time a correct process does so is upper bounded by $\gamma$;
    \item correct processes reveal numbers (adding them to their local history) at the same rate $\lambda$;
    \item at any correct process and within any $\delta$ interval of time, there is a lower bound $th_c$ on the fraction of numbers revealed that come from correct processes.
\end{itemize}

Let $d_1$ and $d_2$ be the selection radius for $V_i$ and $W_i$ respectively (Equation~\ref{eq:witness-critertion}), i.e., a node $j$ is selected as witness if $y_j$ is at distance at most $d_2$ from the origin in $\mathbb{R}_r^b$.

\begin{theorem}[Witness Inclusion]
    As long as $d_1 - d_2 \geq 2 \lambda \gamma$, for any pair of correct nodes $p_i$ and $p_j$, and instance $(\Id,\seq)$: $W_i(\Id,\seq) \subseteq V_j(\Id,\seq)$.
\end{theorem}

In Section~\ref{sec:optimistic} and Appendix~\ref{ap:recovery}, we show how to avoid relying on the above condition by introducing a fallback protocol that ensures progress for correct processes, even in the absence of network synchrony.
Now let $w$ to be the predetermined average witness set size.

\begin{theorem}[Unpredictability]
\label{th:unpredictability}
    Let $S_i$ be the history of process $p_i$ at a particular moment in the execution.
    For any process $p_j$:
    \[\Pr(p_j \in W^{l+\delta}(\cdot,\cdot) \ | \ |S_i| = l) \leq \frac{1}{w} + e^{-\frac{\pi^2 \cdot \delta \cdot th_c}{2br^2}}\]
\end{theorem}

According to our model, it takes $\Delta$ instances for a node corruption to take effect.
In our approach we set $\Delta = 2\delta$ since the same witnesses used in $\WBB$ are involved in the reveal phase $\delta$ instances later.
The the parameters $\delta$, $r$ and $b$ are chosen based depend on the expected speed of adversarial corruption and the probability that the adversary can accurately predict the processes selected as witnesses.
As shown in Section~\ref{subsec:security}, this probability is negligible given a realistic adversarial corruption speed.

\section{Security and Performance}
\label{sec:security}
The witness set selections depend on the positions of the random walks in $\mathbb{R}_r^b$, and there may happen that not enough good nodes or too many bad nodes arrive in the selection zone.
Thus the \emph{average number of instances} until some property of $\WBB$ is violated depends on the random walks properties.
Next, we determine the expected failure time of the complete protocol by calculating the time it takes for the random walks to reach a position where a ``bad'' witness set is chosen.
Note that in our approach, the probability of failure for a particular instance is \emph{dependent} on previous instances, as the random walks depend on their previous state. 

\subsection{Expected Time of Failure}
\label{subsec:security}
\begin{figure}[h]
  \caption{Average number of instances for gathering according to expected witness set size. The parameters used are: $n=1024$, $b = 4$, $r = 2^{10}$, $k = 0.45w$. Each curve is for distinct $t$, the fraction of malicious nodes.}
  \centerline{\includegraphics[width=0.6\textwidth]{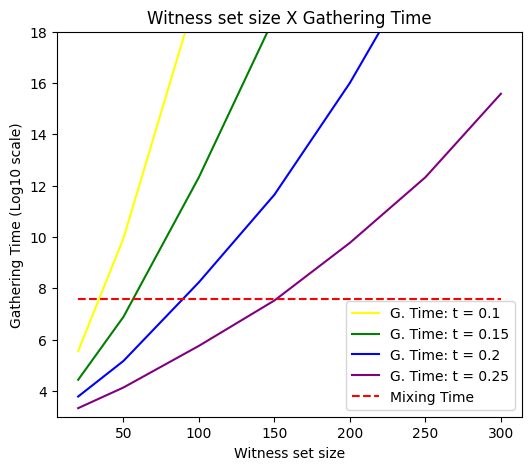}}
  \label{fig:gtMain}
\end{figure}

We say that the long-lived protocol \emph{fails} when, in any $\WBB$ instance, a correct process selects a witness set containing at least $k$ corrupted parties (consistency failure) or fewer than $k$ correct ones (liveness failure).
The adversary may attempt to corrupt processes dynamically to cause protocol failure as quickly as possible. However, as shown in Theorem~\ref{th:unpredictability}, the probability of accurately predicting which processes to corrupt decreases with $\delta \cdot th_c$.
We choose $\delta$ (mixing time in Figure~\ref{fig:gtMain}) to ensure that the \emph{total variation distance} between the $\hashdef$ outcome and the uniform distribution is bounded by $\epsilon < 2^{-20}$, giving the adversary a \emph{negligible advantage} in predicting witnesses.

Given that the $\hashdef$ probability distribution is close to uniform from the time a value is shared to the moment it is added to the history, we consider every step to be independent from the $\hashdef$ states.
In addition, since we are in the random model for hash functions, the distribution of the values issued by malicious nodes becomes uniform for a very large number of values.
We therefore assume that every value added to a local history comprises a random step in $\mathbb{R}_r^b$.

Random walks on $\mathbb{R}_b^r$ may randomly result in the selection of a ``bad'' witness set.
We analyse this scenario bellow.
Since the adversary cannot predict witness sets with significant advantage, we assume all adversarial nodes in the execution are corrupted from start, thereby maximizing the probability of selecting a sufficiently corrupted witness set.

\myparagraph{Gathering Time.} We call \emph{gathering time} the average number of steps until at least $k$ nodes out of $f$ are selected as witnesses, assuming that the initial distribution for each node is uniform and independent.
We give a detailed discussion of the gathering time calculations in Appendix~\ref{ap:passive}.
Intuitively, we relate the problem to the well known \emph{hitting time}~\cite{rand-walk-coder} of Markov chains, in which one calculates the first time the chain reaches a particular set of states.
Let $X_l$ be the number of corrupted nodes in the witness selection area after $l$ steps, we consider a Markov Chain $\{ X_l \ | \ l \in \mathbb{N} \}$ whose possible states are $0,\ldots,f$.
The gathering time is then the average number of steps $l$ such that $X_l \geq k$.
We compare the numbers obtained from our analysis with simulations of $f$ random walks\footnote{We choose $f = 0.25n$, based on the total number of nodes $n$ we consider for each run.} in $\mathbb{Z}_r^b$ in Appendix~\ref{ap:passive}, the results show that our approach gives \emph{conservative}\footnote{In general, our estimations are at least $100$ times smaller than the simulated values.} gathering times.

\myparagraph{Liveness.} The problem is the same when it comes to calculating the time until the witness set is not live.
The possible states are $0,\ldots,n-f$ and we calculate the average number of steps $l$ such that $X_l < k$.

\myparagraph{Selecting the right parameters.} The choice of parameters for the witness oracle depends on the characteristics of the application.
Selecting a bigger diameter $r$ increases the average time of failure, but also increases the mixing time, and thus the system can tolerate adversaries that take longer to corrupt nodes.
We consider the example of a system with $1024$ nodes, and give parameters so that the mixing time is in a practical range.\footnote{The mixing time is $10^{7.56}$ steps for $b = 4$ and $r = 2^{10}$.
If we consider a throughput of $1000$ broadcast messages per second, then the system takes around 10 hours to mix.}
Figure~\ref{fig:gtMain} shows the expected time of failure and the mixing time for $b = 4$, $r = 2^{10}$ and $k = 0.45w$, where $w$ is the expected witness set size.

\myparagraph{Comparison with other probabilistic protocols.}
To contextualize Figure~\ref{fig:gtMain}, we compare our protocol's expected time of failure with two well-known probabilistic protocols: the gossip based reliable broadcast by Guerraoui et al.~\cite{scalable-rb-disc} and the Algorand protocol~\cite{gilad2017algorand}.

Guerraoui et al.~\cite{scalable-rb-disc} propose a one-shot probabilistic reliable broadcast that can be used in Algorithm~\ref{alg:longBroad} to implement $\LPRB$.
In their construction, each instance has an independent probability of failure based on the \emph{sample size},
with a communication complexity of $O(n \cdot \textit{sample size})$, similar to $\WBB$'s $O(n \cdot v)$.
For a system with $1024$ nodes and $t=0.15$, they require a sample size of approximately $250$ nodes to achieve an expected failure time of $10^{12}$ broadcast instances.
In contrast, our protocol achieves the same expected time of failure with just $100$ witnesses.

Algorand~\cite{gilad2017algorand} employs a randomly selected committee to solve Byzantine Agreement,
where the committee size influences the failure probability.
In a system with $t = 0.2$, Algorand requires a committee size of $2000$ nodes to achieve an average failure time of $5 \cdot 10^{9}$, whereas our approach needs fewer than $130$ witnesses.

\myparagraph{Parallel witness sets.} Instead of relying on a single, evolving witness set to validate all messages, multiple independent histories can be maintained, each assigned based on the issuer's ID and sequence number.
This creates parallel dynamic witness sets, where each set validates an equal portion of the workload.
The hitting time for multiple random walks decreases linearly with the number of walks~\cite{alon2008many}, as does the gathering time. However, since each history receives a fraction of the random numbers proportional to the number of witness sets, the total gathering time remains roughly the same.
On the downside, the overall number of delivered messages required to ensure unpredictability increases proportionally with the number of witness sets.

\subsection{Scalability - a Comparative Analysis}
\label{subsec:simulations}

We use simulations to make a comparative analysis between Bracha's reliable broadcast, the probabilistic reliable broadcast from~\cite{scalable-rb-disc} (hereby called \emph{scalable} broadcast) and our witness-based reliable broadcast.
We implemented all three protocols in Golang.
For the simulation software, we used \emph{Mininet}~\cite{lantz2010network} to run all processes in a single machine.

\myparagraph{Setup.} We used a Linode dedicated CPU virtual machine \cite{linode-vm} with $64$ cores, $512$GB of RAM, running Linux 5.4.0-148-generic Kernel \hide{Ubuntu 20.04 OS}with Mininet version 2.3.0.dev6 and Open vSwitch \cite{open-vswitch} version 2.13.8.

\myparagraph{Protocol parameters.} We chose witness set and sample sizes so that both $\WBB$ and scalable broadcast have longevity of $10^6$ instances when $f = 0.1$.
For the witness set size, we use $W = 2\log(n)$ and $V = 3\log(n)$,
the parameters selected for the $\hashdef$ construction are the same as in Figure~\ref{fig:gtMain} and $8$ parallel witness sets are used.
For the sample size we use $5\log(n)$.
The small sizes allow us to better compare the performance of both protocols with Bracha's broadcast for a small number of nodes.

\myparagraph{Description.} Since in a simulated environment we can change network parameters to modify the system's performance, we analyze normalized values instead of absolute ones, using Bracha's protocol as the base line.\footnote{Simulating hundreds of nodes on a single server presents significant performance challenges. In throughput tests involving a large number of nodes, we allocate reduced resources, such as limited bandwidth and CPU time, to each node.} Processes are evenly distributed in a tree topology structure, with $16$ leaves on the base connected by a single parent node.
The bandwidth speed is limited to $20$Mbps per link.

\begin{figure}
    \centering
     \begin{subfigure}[h]{0.48\columnwidth}         \includegraphics[width=1\textwidth]{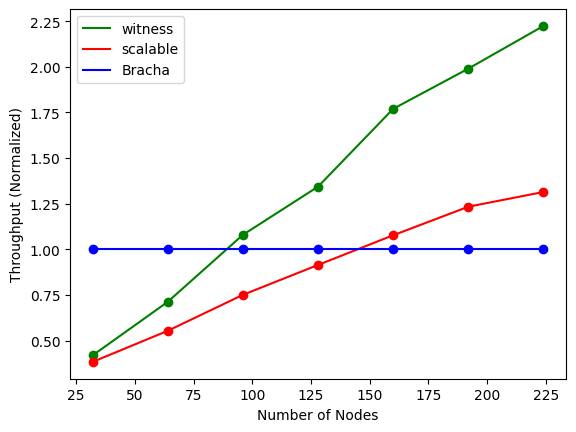}
         \caption{Throughput according to the number of nodes.}
         \label{fig:throughput}
     \end{subfigure}
     \hfill
     \begin{subfigure}[h]{0.48\columnwidth}         \includegraphics[width=1\textwidth]{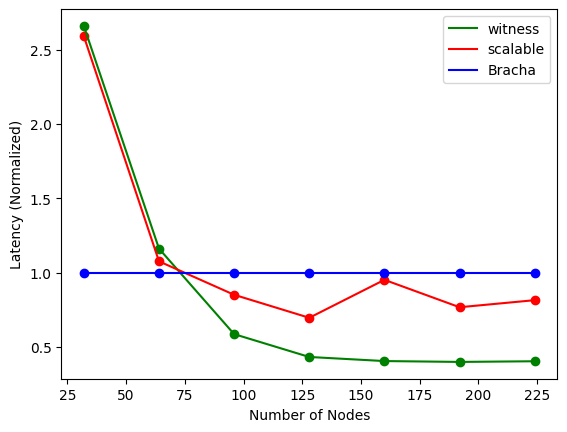}
         \caption{Latency according to the number of nodes.}
         \label{fig:latency}
     \end{subfigure}
     \caption{}
\end{figure}

In the first simulation, we observe each protocol's performance with a high volume of transmitted messages and how it relates to the number of processes. Each process initiates broadcast of a new message once the previous one was delivered throughout all the experiment.
We then measure the achievable throughput (number of delivered messages per second) and average latency for different system sizes as it's show in Figures~\ref{fig:throughput} and~\ref{fig:latency}.

\myparagraph{Comparing protocols.}
The superior asymptotic complexity of the probabilistic protocols is illustrated in Figures~\ref{fig:throughput} and~\ref{fig:latency}, where both $\WBB$ and scalable broadcast show improved performance relative to Bracha’s broadcast as the number of processes increases.
Additionally, $\WBB$ scales consistently better than scalable broadcast, as it requires fewer processes for message validation overall.

\section{Timeout and Recovery}
\label{sec:optimistic}
Based on the assumptions outlined in Section~\ref{subsec:oracleCorrect},
the level of synchrony may contribute to discrepancies in local histories among correct processes.
A process with a significantly divergent history from the rest of the system may attempt to validate messages using a non-responsive witness set.

In this section, we present an extension of our protocol that includes a liveness fallback:
if it takes too long for a process to validate a message, a \emph{timeout} mechanism and a recovery protocol are triggered to ensure progress.

\myparagraph{Recovery Protocol.} First, to account for time, we modify the $\WBB$ initialization block: 
after sampling the witnesses we call the method $\SetTimeout()$(Algorithm~\ref{alg:updates}).
This method guarantees that a $\Timeout$ is triggered after a predetermined amount of time has passed without a successful validation.
Once a process receives a protocol message containing a new broadcast message $m$, it initializes a new instance of message validation and relays $m$ to the corresponding witnesses in a $\NOTIFY$ message.

\begin{algorithm}
\caption{$\WBB$ - Initialization Update}
\label{alg:updates}
\BlankLine

\nl$\rPi$ \Guard{init - $(\Id,\seq)$}{
    \nl $V_i \gets \omega_i.\GetAllWitnesses(\Id,\seq)$\;
    \nl $W_i \gets \omega_i.\GetOwnWitnesses(\Id,\seq)$\;
    \nl $\SetTimeout()$\;
}
\BlankLine

\end{algorithm}

When a $\Timeout$ event is triggered, a process resorts to the recovery protocol outlined in Algorithm~\ref{alg:recovery} (Appendix~\ref{ap:recovery}).
The actions on the recovery protocol should be executed by every process, and are performed \emph{only once a process triggers $\Timeout$ or delivers a message in that instance}.\footnote{Messages from the recovery protocol received by a process before satisfying any of these conditions are then stored and treated later.}

The idea behind the algorithm is to ``recover'' any value that could have possibly been delivered in the witness validation procedure, and then execute Bracha's Byzantine reliable broadcast~\cite{bra87asynchronous}.
We achieve this by making processes echo the last $\WBB$ they acknowledged, so that if a message is delivered by any process using $\WBB$, no distinct message can be delivered using the recovery protocol.
A detailed description of the protocol, as well as correctness proofs can be found in Appendix~\ref{ap:recovery}.

The inclusion of the recovery protocol ensures progress if $\WBB$ is not live for a correct process, albeit with a higher communication cost as we use Bracha's Byzantine reliable broadcast which has communication complexity of $O(n^2)$.
Reduced communication using $\WBB$ is thus achieved in \emph{optimistic} runs, when network asynchrony is not too severe.

\section{Related Work}\label{sec:relwork}

Solutions designed for partially synchronous models~\cite{DLS88} are inherently \emph{optimistic}: safety properties of consensus are always preserved, but liveness is only guaranteed in sufficiently long periods of synchrony, when message delays do not exceed a pre-defined bound.
This kind of algorithms, also known as \emph{indulgent}~\cite{indulgent}, were originally intended to solve the fundamental consensus problem, which gave rise to prominent partially synchronous state-machine replication protocols~\cite{paxos,pbft} and, more recently, partially-synchronous blockchains.

Reliable broadcast protocols~\cite{bra87asynchronous} provide a weaker form synchronization than consensus: instead of reaching agreement on the total order of events, reliable broadcast only establishes common order on the messages issued by any given source.      
This partial order turns out handy in implementing ``consensus-free'' \emph{asset transfer}~\cite{at2-cons,Gup16}, by far the most popular blockchain application, and resulting implementations are simpler, more efficient and more robust than consensus-based solutions~\cite{fastpay,astro-dsn,pastro21disc}.\footnote{These implementations, however, assume that no account can be concurrently debited. i.e., no conflicting transactions must ever be issued by honest account owners~\cite{crypto-concurrency}.}
To alleviate $O(n^2)$ communication complexity of classical \emph{quorum-based} reliable broadcast algorithms~\cite{br85acb,ma97secure,MR97srm,toueg-secure}, one can resort to \emph{probabilistic} relaxations of its properties~\cite{probabilisticquorums,scalable-rb-disc}.
The probabilistic broadcast protocol in~\cite{scalable-rb-disc} achieves, with a very high level of security, $O(n \log n)$ expected complexity and $O(\log{n}/ \log\log{n})$ latency by replacing quorums with randomly selected samples of $O(\log n)$ size.
In this paper, we propose an optimistic probabilistic reliable broadcast algorithm that, in a good run, exhibits even better security (due to the ``quasi-deterministic'' though unpredictable choice of witnesses) and, while achieving the same communication complexity, constant expected latency. 
For simplicty, as a fall-back solution in a bad run, we propose the classical $O(n^2)$ broadcast protocol~\cite{bra87asynchronous}. 
However, one can also use the protocol of~\cite{scalable-rb-disc} here, which would give lower costs with gracefully improved security.   

In the Algorand blockchain~\cite{gilad2017algorand,chen2016algorand}, scalable performance is achieved by electing a small-size \emph{committee} for each new block. 
To protect the committee members from a computationally bound adversary, one can use verifiable random functions (VRF)~\cite{micali1999verifiable}. 
The participants use a hash of the last block and their private keys as inputs to the VRF that returns a \emph{proof of selection}.
As the proof is revealed only when a committee member proposes the next block, the protocol is protected against an adaptive adversary.
The approach was later applied to asynchronous (randomized) Byzantine consensus~\cite{cohen2020not}, assuming trusted setup and \emph{public-key infrastructure} (PKI).
Similar to~\cite{gilad2017algorand,cohen2020not}, we use local knowledge for generating unpredictable process subsets of a fixed expected size.
In contrast, our protocol does not assume trusted setup. We generate pseudo-randomness based on the local states, without relying on external sources (e.g., the blockchain state). 
However, we only tolerate a \emph{slow} adversary (time to corrupt a node considerably exceeds communication delay).

\section{Conclusion}
\label{sec:conclusion}

In this work, we introduced a new probabilistic broadcast protocol aimed at enhancing the performance and scalability of decentralized systems, even under a slow adaptive adversary.
By utilizing a small number of witnesses for message validation, $\WBB$ achieves consistent performance improvements compared to other approaches when the system size is large.

Our approach provides flexibility by allowing key parameters to be fine-tuned for optimizing performance under varying network conditions.
This adaptability makes the protocol well-suited for a range of real-world applications where network characteristics, such as latency and workload, may fluctuate.
Future work could further explore how this flexibility can be leveraged by integrating dynamic membership mechanisms into the protocol.
For instance, in the context of asset transfer systems, witness selection could be adapted to reflect the stake or reputation each node holds within the network, allowing more influential or trusted participants to take on a greater role in the validation process.

\bibliographystyle{plainurl}
\bibliography{main.bbl}
%
\appendix

\section{Secret Sharing Scheme}
\label{app:secret}
In order to prevent the adversary from biasing the outcome of $\hashdef$, we require each number to be generated by a verifiable source of randomness.
This is achieved using the signature scheme and a hash function.
Consider a random string $s$ known to all nodes, to generate the random number $x$ of instance $(p_i,\seq)$, $p_i$ signs $s \oplus (p_i,\seq)$ and assigns to $x$ the hash of the resulting signature.
The signature for $s \oplus (p_i,\seq)$ is then used as proof that $x$ was correctly generated.
The random string $s$ used for $(p_i,\seq)$ is the random number $x'$ generated in the previous instance, and the original seed can be any agreed upon number.

On top of that, we capitalize on the steady distribution of a random walk
which is the uniform distribution in a torus\footnote{To guarantee uniform distribution, the diameter $r$ needs to be odd~\cite{berestycki2016mixing}, we can trivially achieve this by skipping the last point (so the diameter of each dimension becomes $r-1$).} (this means that as we increase the number of steps, the distribution of $\hashdef$ converges to uniform).
The number of steps necessary to make the distribution of the random walk close to uniform is called the \emph{mixing time}.
For each new random number that is broadcast by a node, we delay its addition to the history by $\delta = mixTime/th_c$ steps, where $th_c$ is the fraction of the throughput generated by correct sources.\footnote{One can adjust $th_c$ based on the broadcast rate of the $n-f$ nodes with smallest rate. Alternatively one can wait until the total number of additions to the history, committed by the  $n-f$ nodes with smallest rate, has reached $mixTime$.} This guarantees that the adversary cannot use it's current state to issue carefully chosen numbers and make the outcome of $\hashdef$ close to any specific value, in other words, the adversary does not know if the delayed step will bring the value of the hash closer of farther to a desirable value.

\myparagraph{Secret Sharing.}
The unpredictability is achieved with a secret sharing protocol, in which the source's signature for $(\Id,\seq)$ is shared and only revealed after $\delta$ messages are delivered.
The goal of the protocol is to guarantee that: 

\begin{itemize}
    \item Correct processes agree on the revealed number.

    \item No information about a number issued by a correct process is revealed until at least one correct process starts the reveal phase.
\end{itemize}

We use Shamir's secret sharing scheme~\cite{shamir1979share} abstracted as follows:
the split method $\SH.\Split(x,n,f+1)$ takes a string $x$ and generates $n$ shares such that any $f+1$ shares are sufficient to recover $x$.
The recover method $\SH.\Recover(X)$ takes a vector with $f+1$ shares as input and outputs a string $x'$ such that, if $X$ was generated using $\SH.\Split(x,n,f+1)$, then $x = x'$.
In addition, no information about $x$ is revealed with $f$ or less shares.

The pseudo-codes in Algorithms~\ref{alg:commitShare} and~\ref{alg:commitReveal} describe the complete protocol.
In short, the source $p$ first generates the signature $\pi$ for $s \oplus (\Id,\seq)$ and splits $\pi$ in $n$ shares.
Next, $p$ includes the corresponding share to each $\NOTIFY$ message.
We assume that the source's signature for the message with the share is also sent alongside it.

\begin{algorithm}
\caption{Secret Sharing Protocol - Sharing Phase}
\label{alg:commitShare}
\BlankLine

\nl$\rS$ \Guard{preparation for broadcasting $m$}{
    \nl $\pi \gets \Sign(s \oplus (\Id,\seq))$\;
    \nl $\overline{\pi} \gets \SH.\Split(\pi, n, f+1)$\;
    \nl Include $\overline{\pi}[i]$ in $\NOTIFY$ message sent to $p_i$\;
}
\BlankLine

$\nl\rPi$ \Operation{$\deliver(m,\Id,\seq)$}{
    \nl $H_i \gets H_i \cup \{m\}$\;
    \nl $\Checkpoint[\Id,\seq] \gets |H_i|$\;
} 
\BlankLine
\end{algorithm}

\begin{algorithm}
\caption{Secret Sharing Protocol - Reveal Phase}
\label{alg:commitReveal}
\BlankLine

\nl$\label{line:checkpoint}\rPi$ \Guard{$|H_i| - \Checkpoint[\Id,\seq] = \delta$}{
    \nl send $\langle \REVEAL, \Id, \seq, \overline{\pi}[i] \rangle$ to every $w \in V_i$
}
\BlankLine

\nl$\rW$ \Guard{receiving $\langle \REVEAL, \Id, \seq, \overline{\pi}[j] \rangle$ from $f+1$ processes}{
    \nl $\pi \gets \SH.\Recover(\overline{\pi}_{f+1})$\Comment*[r]{$\overline{\pi}_{f+1}$ is a vector with $f+1$ shares}
    \nl \If{$\Verify(\pi, s \oplus (\Id,\seq),pk_{\Id})$} {
        \nl send $\langle \DONE, \pi, \Id, \seq \rangle$ to every $p \in \Pi$\;
    }
    \nl \Else{
        \nl Let $S$ be the set of $f+1$ signed shares\;
        \nl send $\langle \FAILED, S, \Id, \seq \rangle$ to every $p \in \Pi$\;
    }
}
\BlankLine

\nl$\rPi$ \Guard{receiving $\langle \DONE, \Id, \seq, \pi \rangle$ from a witness}{
    \nl \label{line:checkHash} \If{$\Verify(\pi, s \oplus (\Id,\seq),pk_{\Id})$ and $|H_i| - \Checkpoint[\Id,\seq] \geq \delta$} {
        \nl \If{$\Secret[\Id][\seq] \neq \top$}{
            \nl $\Secret[\Id][\seq] \gets h(\pi)$\;
            \nl send $\langle \DONE, \Id, \seq, \pi \rangle$ to every $w \in V_i$\; \label{line:relayDone}
        }
    }
}
\BlankLine

\nl$\rPi$ \Guard{receiving $\langle \FAILED, S, \Id, \seq \rangle$ from a witness}{
    \nl $\Convicted \gets \Convicted \cup \{\Id\}$\;
    \nl send $\langle \FAILED, S, \Id, \seq \rangle$ to every $w \in V_i$\;
}
\BlankLine

\nl$\rW$ \Guard{receiving $\langle \DONE, \Id, \seq, \pi \rangle$}{
    \nl \If{$\Verify(\pi,s \oplus (\Id,\seq),pk_{\Id})$}{
        \nl send $\langle \DONE, \Id, \seq, \pi \rangle$ to every $p \in \Pi$\;
    }
}
\BlankLine

\nl$\rW$ \Guard{receiving $\langle \FAILED, S, \Id, \seq \rangle$}{
    \nl send $\langle \FAILED, S, \Id, \seq \rangle$ to every $p \in \Pi$\;
}
\BlankLine
\end{algorithm}

When delivering the message associated with $(\Id,\seq)$, nodes store the current number of delivered messages in the history.
After delivering $\delta$ new messages, each process starts executing the reveal phase using the same witnesses to recover the secret.
Each process then sends its share to the witnesses which, after gathering enough shares, either reveal the secret or build a proof that the shares were not correctly distributed by the source.
In the former case, each process verifies the revealed signature $\pi$ and adds the hash of $\pi$ to $\Secret$, defining a step in the random walk on $\hashdef$.
In the latter, when receiving such proof, a process marks the faulty source as ``convicted''.

The following results assume correctness of the underlying $\WBB$ protocol.

\begin{proposition}
    If $p_i$ and $p_j$ correct add $x$ and $x'$ to $\Secret[\Id,\seq]$ respectively, then $x = x'$.
\end{proposition}

\begin{proof}
    Before adding $x$ or $x'$ to $\Secret$, they check whether the revealed signature $\pi$ is valid for the string $s \oplus (\Id,\seq)$ (line~\ref{line:checkHash}).
    From the uniqueness property of the signature scheme, it follows that $x = x'$.
\end{proof}

\begin{proposition}
    If a correct process $p_i$ adds $x$ to $\Secret[\Id,\seq]$, then eventually every correct process adds $x$ to $\Secret[\Id,\seq]$.
\end{proposition}

\begin{proof}
    After receiving $\langle \DONE, \Id, \seq, \pi \rangle$ from a witness, $p_i$ relays the message to every witness in $V_i$ (line~\ref{line:relayDone}).
    Because the underlying $\WBB$ is correct, $W_j \subseteq V_i$ for any $p_j$ correct.
    Thus, there is a common correct witness that receives $\pi$ from $p_i$ and relays it to $p_j$, which adds $h(\pi)$ to $\Secret$ (note that since $p_i$ added $x = h(\pi)$ to secret, $\pi$ is a valid signature).
\end{proof}

\begin{proposition}
    Let $\pi$ be the signature a correct process shares at instance $(\Id,\seq)$, then any other process can only reveal $\pi$ if at least one correct process delivers $\delta$ new messages after delivering $(\cdot, \Id, \seq)$.
\end{proposition}

\begin{proof}
    From Shamir's construction, no information about $\pi$ is revealed unless a process gathers $f+1$ shares~\cite{shamir1979share}.
    A correct process does not reveal its share unless it has delivered $\delta$ new messages (line~\ref{line:checkpoint}), and a correct source does not reveal the secret $\pi$.
    It follows that $\pi$ can be revealed only when at least one correct process reveals its share.
\end{proof}

\myparagraph{Incorrect distribution of shares.}
Detecting when a source incorrectly distributes shares is important to prevent the adversary from controlling when the secret is revealed.
A malicious process can send malformed shares that cannot be recovered by correct processes, and later (in an arbitrary time) send $\DONE$ with the correct secret through a corrupted witness.
We employ the detection mechanism to discourage malicious sources from distributing bad shares.
Processes relay the proof as soon as an incorrect share is detected, leaving little time for malicious nodes to take advantage of the incorrect distribution.
Correct processes can subsequently exclude the misbehaving party from the system.

One can also guarantees the recovery of a secret (independently of adversarial action) with a stronger primitive called \emph{Verifiable Secret Sharing} (VSS~\cite{das2021asynchronous}).
The asynchronous VSS protocol presented in~\cite{das2021asynchronous} allows nodes to check their shares against a commitment that must be reliably broadcast, thus preventing a malicious source from distributing bad ones.
One can replace the reliable broadcast with $\WBB$, attaching the commitment to the broadcast message to reduce communication complexity.
Because the size of the commitment can be large and the scheme computationally intensive, we use Algorithms~\ref{alg:commitShare} and~\ref{alg:commitReveal} as a more efficient approach.
For the remaining of the paper, we consider \emph{optimistic} runs of the algorithms in which the detection mechanism successfully deter malicious participants from distributing incorrect shares.

\section{Witness Oracle - Correctness}
\label{app:oracleCorrect}

The pseudo-code for the witness oracle implementation is described in Algorithm~\ref{alg:witOracle}.
Parameters $d_1$ and $d_2$ are the distances of the selection area for $V_i$ and $W_i$ respectively, while $S_i$ is node $i$'s current local history of random numbers.

\begin{algorithm}
\caption{Local Witness Oracle - code for process $i$.}
\label{alg:witOracle}
\BlankLine

\textbf{Local Variables:} \\
$W_i, V_i \gets \emptyset$\;
\BlankLine

\nl \Operation{$\DefineWitness(\Id,\seq)$}{
    \nl \For{$j \in \Pi$}{
        \nl $y_j \gets f_j(S_i)$\;
        \nl \If{$TorusDist_{r,b}(y_j,[0]) \leq d_1$}{
            \nl $V_i[\Id,\seq] \gets V_i[\Id,\seq] \cup \{j\}$\;
        }
        \nl \If{$TorusDist_{r,b}(y_j,[0]) \leq d_2$}{
            \nl $W_i[\Id,\seq] \gets W_i[\Id,\seq] \cup \{j\}$\;
        }
    }
}
\BlankLine

\nl \Operation{$\GetAllWitnesses(\Id,\seq)$}{
    \nl \If{$V_i[\Id,\seq] = \emptyset$}{
        \nl $\DefineWitness(\Id,\seq)$\;
    }
    \nl \textbf{return} $V_i$\;
}
\BlankLine

\nl \Operation{$\GetOwnWitnesses(\Id,\seq)$}{
    \nl \If{$W_i[\Id,\seq] = \emptyset$}{
        \nl $\DefineWitness(\Id,\seq)$\;
    }
    \nl \textbf{return} $W_i$\;
}
\BlankLine
\end{algorithm}

We now analyse the properties of the resulting protocol composed of Algorithms~\ref{alg:longBroad},~\ref{alg:witOracle},~\ref{alg:commitShare} and~\ref{alg:commitReveal}.
To combine Algorithm~\ref{alg:witOracle} with~\ref{alg:commitReveal}, we make $S_i \equiv \bigcup{\Secret[\Id][\seq]}$.
First, to verify the conditions under which \textit{Witness Inclusion} and \textit{Unpredictability} are satisfied, we assume that:

\begin{itemize}
    \item The interval between the first time a correct process adds $x'$ to $\Secret$, and the last time a correct process does so is upper bounded by $\gamma$;
    \item correct processes add new numbers to $\Secret$ at the same rate $\lambda$;
    \item at any correct process and within any $\delta$ interval of time, there is a lower bound $th_c$ on the fraction of numbers added to $\Secret$ that come from correct processes.
\end{itemize}

Recall that $d_1$ and $d_2$ in Algorithm~\ref{alg:witOracle} are the selection radius for $V_i$ and $W_i$ respectively, i.e., a node $j$ is selected as witness if $y_j$ is at distance at most $d_2$ from the origin in $\mathbb{R}_r^b$.

\setcounter{theorem}{3}
\begin{theorem}[Witness Inclusion]
    As long as $d_1 - d_2 \geq 2 \lambda \gamma$, for any pair of correct nodes $p_i$ and $p_j$, and instance $(\Id,\seq)$: $W_i(\Id,\seq) \subseteq V_j(\Id,\seq)$.
\end{theorem}

\begin{proof}
    Let $R_i(t)$ and $R_j(t)$ be $p_i$'s and $p_j$'s histories at a particular time $t$ respectively.
    The condition $W_i(\Id,\seq) \subseteq V_j(\Id,\seq)$ is guaranteed if $TorusDist_{r,b}(f_i(R_i),f_i(R_j)) \leq d_1 - d_2$.

    By assumption, all numbers in $R_i(t-\gamma)$ are already in $R_j(t)$.
    The only numbers that might be in $R_i(t) \cup R_j(t) - R_i(t) \cap R_j(t)$ are those $p_i$ and $p_j$ have added to their histories in $(t-\gamma, t]$, therefore, because they add numbers at a maximum rate of $\lambda$:
    \[|R_i(t) \cup R_j(t) - R_i(t) \cap R_j(t)| \leq 2 \lambda \gamma\]
    Moreover, $TorusDist_{r,b}(f(S),f(T)) \leq |S \cup T - S \cap T|$ from the \emph{Locality} property of $\hashdef$. Thus $W_i(\Id,\seq) \subseteq V_j(\Id,\seq)$ is satisfied whenever $d_1 - d_2 \geq 2 \lambda \gamma$.
\end{proof}

\myparagraph{Mixing Time.} We generalize an upper bound on the mixing time of a random walk on the circle $\mathbb{Z}_n$~\cite{berestycki2016mixing},
by including a multiplicative factor of $b$ to calculate the mixing time of a random walk on $\mathbb{Z}_r^b$:
\[ \MixTime(\mathbb{Z}_r^b, \epsilon) \leq  \frac{-2b \cdot r^2 \cdot ln(\epsilon)}{\pi^2}\]

Where $\epsilon$ is the upper bound on the distance between the random walk distribution $\mu$ and the uniform distribution $\upsilon$,
using the \emph{total variation distance}:
\[ \TotalVar(\mu,\upsilon) = \frac{1}{2}\sum_{x \in \mathbb{Z}_r^b} |\mu(x) - \upsilon(x)| < \epsilon\]

Consider $w$ to be the predetermined average witness set size, and $d$ the corresponding distance selected to achieve $w$.

\begin{theorem}[Unpredictability]
    Let $S_i$ be the history of process $p_i$ at a particular moment in the execution.
    For any process $p_j$:
    \[\Pr(p_j \in W^{l+\delta}(\cdot,\cdot) \ | \ |S_i| = l) \leq \frac{1}{w} + e^{-\frac{\pi^2 \cdot \delta \cdot th_c}{2br^2}}\]
\end{theorem}

\begin{proof}
    Let $y_j = f_j(S_i)$, $S_i^{\delta}$ be $p_i$'s history after adding $\delta$ new numbers to $S_i$ and $y_j' = f_j(S_i^{\delta})$.
    Calculating the probability that $p_j \in W^{l+\delta}(\cdot,\cdot)$ is equivalent to calculate the probability that $y_j'$ is within distance $d$ from the origin.

    Any of the next $\delta$ numbers $p_i$ will reveal must be already shared by its source and the corresponding instance finished delivered by $p_i$, so that $p_i$ can start counting the number of instances that passed before revealing the number.
    Any number originated from a correct process is shared regardless of the current state and thus comprise a random step in $\mathbb{R}_r^b$ when applying $f_j$, note that there are at least $\delta \cdot th_c$ such numbers.
    On the other hand, numbers originated from malicious nodes were already shared based on information up to $S_i$.
    Thus, we can apply these numbers in any order from $S_i$, since they do not depend in later states of $p_i$'s history.

    Consider the resulting hash value $y_j''$ after applying all numbers issued by malicious nodes at once from $S_i$.
    Now, when we apply the remaining steps coming from correct processes, it is equivalent of performing a random walk starting from $y_j''$, and the upper bound on the mixing time applies to this case as well.

    The probability that $y_j'$ is within distance $d$ from the origin is given by the resulting distribution $\upsilon$ after applying all remaining steps.
    If the distribution is uniform, then the probability is $1/w$.
    The maximum the probability can variate is given by the $\TotalVar$, which when replacing $\epsilon$ becomes:
    \[\TotalVar(\mu,\upsilon) < e^{-\frac{\pi^2 \cdot N_{steps}}{2br^2}}\]
    Where $N_{steps}$ is the number of steps in the random walk.
    The result follows by replacing $N_{steps}$ with the minimum amount of numbers coming from correct processes: $\delta \cdot th_c$.
\end{proof}

\section{Security Against Passive Attacks}\label{ap:passive}
In the passive attack, the adversary waits until the history hash will be such that the selected witness set for a malicious message will contain many compromised nodes. The selection of each compromised node depends on what can be considered as an independent random walk (the per node history hash) arriving to a small subspace in $\mathbb{Z}_r^b$ (within a defined distance from the origin). 

The expected ratio of compromised nodes in the witness set is their ratio in the total population. We argue that the number of random walk steps (new numbers revealed and added to the local history) required to obtain a much higher ratio of compromised witnesses can be very high. One key indication that it takes long\hide{for low likelihood} for multiple random walks to co-exist in the same region in $\mathbb{Z}_r^b$ is the linear speedup in parallel coverage time of $\mathbb{Z}_r^b$ \cite{rand-walk-coder}.

\subsection{The Gathering Time Problem}
We are interested in the average occurrence time of two events: the time until the number of selected compromised nodes exceeds a given ratio of the expected witness set size, and the time until the number of selected correct nodes is smaller than said ratio. The former event is discussed first, which we call \emph{Gathering Time} since the adversary waits until enough random walks gather in a defined area.

Our problem is related to Hitting Time bounds~\cite{rand-walk-coder} that were well studied in recent years  and consider the time it takes for one (or many) random walks to arrive to a specific destination point.
The Gathering Time differs from the Hitting Time in two main aspects: 1) we require that the random walks will be at the destination area at the same time and 2) we do not require all random walks to arrive at the destination but instead we are interested in the first time that a subset of them, of a given size, will arrive at the destination area. Next we provide an approximation for a lower bound which we compare to simulated random walks results.

\subsection{A simplified Markov Chain approach}

We assume that nodes are initially mapped to points in $\mathbb{Z}_r^b$ and are selected to be witnesses if they are at distance at most $r \cdot q / 2$ from the origin in $\mathbb{Z}_r^b$ (based on $L_{\infty}$).
The initial mapping of nodes is uniform and independent, also, the movement of nodes in the space can me modeled as independent random walks.

The initial probability that a node $i$ is found at distance at most $p\cdot r$ is equal to the probability that in all $b$ dimensions its location is at most $p\cdot r$, i.e. $Pr(dist_i\leq p\cdot r)=(2p)^b$.
Therefore, $q^b$ is the initial probability that a node is selected and the expected witness set size is $n \cdot q^b$.
We consider configurations where the witness set size is logarithmic with $n$, i.e., $n\cdot q^b = c \log_2 n$.

Next, we estimate the expected time until at least $k$ out of $f$ compromised nodes will be selected, where $t < n/3$ and $s > t$ are the ratio of compromised nodes in the system and in the witness set respectively, i.e., $f = t \cdot n$ and $k = s \cdot c \cdot n \cdot q^b = s \cdot c\log_2n$.
Initially, $f$ nodes are uniformly distributed in $\mathbb{Z}_r^b$.
In one step, every node moves a single unit ($+1$ or $-1$) in one dimension.

\myparagraph{Simplification idea.} 
Let $A$ be the witness selection area in $\mathbb{Z}_r^b$ and $B = \mathbb{Z}_r^b - A$, that is, $B$ is the complement of $A$ in $\mathbb{Z}_r^b$.
Calculating the hitting time for $k$ out of $f$ nodes to reach $A$ is an analytical and computational challenge.
Suppose, for instance, that we use the transition matrix $\mathbb{P}$ of a random walk in $\mathbb{Z}_r^b$, depending on the size of the space, the number of states makes it implausible to solve the hitting time computationally using $\mathbb{P}$ (even without considering the $f$ simultaneous random walks).

Instead, we represent the number of nodes inside $A$ as a Markov Chain $\{X_l \ | \ l \in \mathbb N \}$, where $S = 0, ..., f$ are the possible states and $X_l$ is the number of nodes in the targeted area after $l$ steps.
In order to calculate $\Pr(X_{l+1} = y \ | \ X_{l} = x)$ we consider, as a simplification, only the average probability (over all possible positions) that each individual node inside $A$ may leave the area in one step, as well as the average probability that each individual node inside $B$ may move to $A$.
These probabilities are assumed to be the same for every node independently of its position inside $A$ (or inside $B$).

Only nodes located in the ``border'' of $A$ with $B$ can move to $B$ in one step.
Intuitively, the border of $A$ is composed of points that are one step away from a point in $B$.
To formalize this notion we use the norm-$1$ distance, which can be thought as the minimum number of steps needed to go from a point $u$ to another point $v$.
\[ \LOneDist_{r,b}(u, v) := \sum_{j=0}^{b-1} \Min(u_j-v_j \mod{r}, v_j-u_j \mod{r}) \]

We define the distance between a point and an area $B \subseteq \mathbb{Z}_r^b$ as the minimum distance from $u$ to any point of $B$.
The border of an area $A$ is comprised of all the points that are at distance $1$ from its complement $B$.
\[ \Border(A) := \{ u \in A \ | \ \LOneDist_{r,b}(u,B) = 1 \} \]

Nodes inside $A$ that are not located in the border cannot leave the area in a single step.
Thus, the average probability that a node leaves $A$ after one step (over all points of $A$\footnote{As shown later, the probability that a node leaves $B$ given that it is in $\Border(B)$ is the same for every point in $\Border(B)$, which is not the case for all points in $\Border(A)$.}) is:
\[ \AVG\Pr(i \ leaves \ A) = \frac{\Pr(i \ leaves \ A \ | \ i \in \Border(A)) \cdot |\Border(A)|}{|A|} \]

The probability of a node leaving $B$ is analogous.

Let $Y_l$ denote the number of nodes that leave $A$ from step $l$ to step $l+1$, and $W_l$ the number of nodes that leave $B$.
We define $C_l = W_l - Y_l$ as the variation of the number of nodes in $A$, thus:
\[ X_{l+1} = X_l + C_l \]
\[ \Pr(X_{l+1} = y \ | \ X_l = x) = \Pr(C_l = y - x \ | \ X_l = x) \]

\myparagraph{Probability calculations.}
The witness selection region $A$ is comprised of all the points at a distance $d = r \cdot q/2$ ($L_{\infty}$) from the origin.
We say that $u = [u_1,...,u_b] \in A$ $\Iff$ $\forall u_i: -d \leq u_i \leq d$ (where $-d$ is the same as $r - d$).
The size of the space (number of points) is $r^b$, while the size of $A$ is $(2d + 1)^b$, and thus $|B| = r^b - (2d + 1)^b$.

A point $v$ inside $B$ that is one step away from $A$ satisfy the following condition: there is a single $v_i$ such that $v_i = d+1$ or $v_i = -(d+1)$, and for all other $v_j$, $-d \leq v_j \leq d$.
For a specific $v_i$, there are $(2d+1)^{d-1}$ points in $\Border(B)$ with $v_i = d+1$ (same for $v_i = -(d+1)$).
Since there are $b$ dimensions, the total number of points in $\Border(B)$ is $2b(2d+1)^{b-1}$.
Now for any node in $\Border(B)$, a random step can change the value of a single position by $+1$ or $-1$, so the probability that such node moves to $A$ after one step is $\frac{1}{2b}$.

On the other hand, in $\Border(A)$ there are points that are at distance $1$ from multiple points in $B$.
A point $u$ in $\Border(A)$ should satisfy: $\forall u_i, -d \leq u_i \leq d$ and $\exists u_j: u_j = d$ or $u_j = -d$.
The probability of moving from $\Border(A)$ to $B$ depends on how many values $u_j = d$ (or $-d$),
and there can be at most $b$ values equal to $d$.
Suppose there are $k$ such $u_j$, then the probability that a node in this point leaves to $B$ is $\frac{1}{2(b-k+1)}$.

There are $\binom{b}{k}$ distinct ways of choosing $k$ out of $b$ values to be either $d$ or $-d$, and for each choice of $k$ elements, there are $2^k$ ways of arranging it in a $k$-sequence of $-d$ and $d$.
Finally, there are $2^k\binom{b}{k}(2d - 1)^{b-k}$ points $u$ in which $k$ $u_i$ have value either $d$ or $-d$.
The total number of points in $\Border(A)$ is the sum over all possible $k$:
\[ |\Border(A)| = \sum_{j=1}^{b}2^j\binom{b}{j}(2d-1)^{b-j} \]

We can now calculate the average probability that a node in $\Border(A)$ moves to $B$:
\[ \AVG\Pr(i \ leaves \ A \ | \ i \in \Border(A)) = \frac{\sum_{j=1}^{b}\frac{2^j\binom{b}{j}(2d-1)^{b-j}}{2(b-j+1)}}{|\Border(A)|} \]

Next, we calculate each component of the transition matrix.

\myparagraph{Transition Matrix.} Let $p_A = \AVG\Pr(i \ leaves \ A)$ and $p_B = \AVG\Pr(i \ leaves \ B)$ the individual probabilities that a node leave the current area in one step.
Suppose that there are $x$ nodes in $A$ and $f-x$ nodes in $B$. Then,
\[ Y_l \sim \Binomial(x,p_A) \]
\[ W_l \sim \Binomial(f-x,p_B) \]

The variation $C_l$ is calculated as following:
\[ \Pr(W_l - Y_l = c \ | \ X_l = x) = \sum_{c_1-c_2 = c}\Pr(W_l = c_1 \ | \ X = x)\Pr(Y_l = c_2 \ | \ X = x) \]

Where $c_1$ can range from $0$ to $f-x$ and $c_2$ can range from $0$ to $x$.
Let $P$ be the transition matrix of the Markov Chain $\{X_l \ | \ l \in \mathbb{N} \}$, and $P_{ij}$ the component of the $i^{th}$ row and $j^{th}$ columns, then:
\[ P_{ij}  = \Pr(X_{l+1} = j \ | \ X_l = i) = \Pr(W_l - Y_l = j - i \ | \ X_l = i)\]

\myparagraph{Hitting Time.} We use $P$ to compute the average number of steps that starting from a particular state $S$, the chain arrives at a state $S' \geq k$.
The average hitting time is the weighted sum of all individual hitting times, where the weight for each initial state is the probability of that state in the initial distribution.
To calculate the initial distribution, we assume that all nodes are distributed uniformly at random and have the same probability $q^b$.
Let $X_{init}$ denote the initial number of nodes inside $A$, then:
\[ X_{init} \sim \Binomial(f, q^b) \]

\myparagraph{Simulations.} We ran simulations for multiple random walks in $\mathbb{Z}_r^b$ and compared the hitting time of the simulations with the values obtained from the simplified analysis.
For this purpose, we chose parameters so that the number of steps required for witness set corruption is small enough to allow the simulations to finish in a reasonable time.

\begin{figure}[htbp]
  \caption{Simulations (thick line) and simplified analysis (dashed line). On the left graph: $r = 2^{10}$. On the right graph: $b = 8$. In both graphs $t = 0.25$, $c = 2$ and $s = 0.5$.}
  \centerline{\includegraphics[width=14cm]{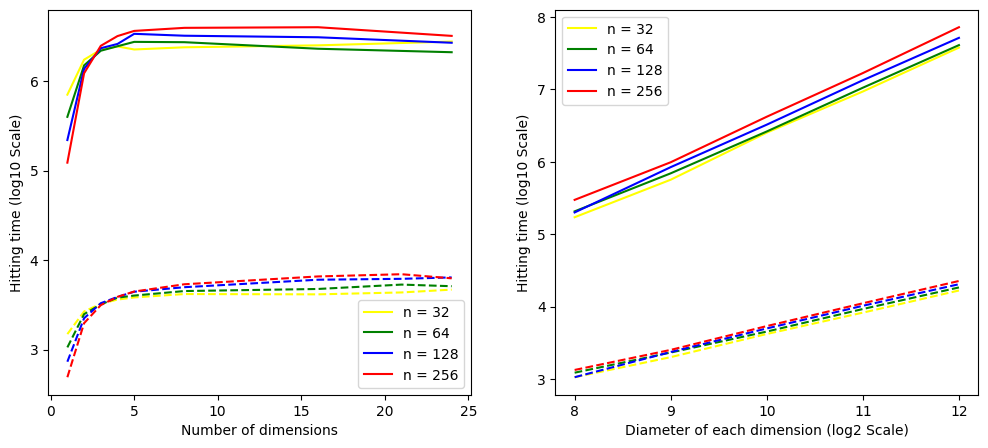}}
  \label{fig:comparativeHit}
\end{figure}

Figure~\ref{fig:comparativeHit} shows results from the simulations and from the simplified analysis.
For each point in the graphs, we took the average of $1000$ runs.
From both graphs, we can conclude that the simplified analysis gives a \emph{conservative} estimate of the hitting time.

\myparagraph{Liveness.} In the counterpart problem of the Gathering Time, we want to know the average number of steps after which there are not enough correct nodes in the witness set.
It can be calculated with simple modifications in the Markov chain: the possible states are now $0,\ldots,n-f$, and we use the updated $P$ to compute the average number of steps until the chain arrive at a state $S' < k$.

\myparagraph{History differences.} Because of history differences, the path of local random walks may slightly diverge among processes.
Consider the extreme case where each local history evolves independently from each other (and thus the difference can be arbitrarily large).
The \emph{speed up} property of multiple random random walks guarantees that a linear speed up (on the number of walks $f$) of the hitting time occurs~\cite{alon2008many}.
In reality, the walks are highly correlated and are apart from each other only by a factor that is much smaller than the hitting time.
Thus, the speed up in the cases we consider is negligible in relation to the hitting time.

\subsection{Liveness vs Consistency graph}

\begin{figure}[h]
  \caption{Liveness and Consistency failure times. The parameters used are: $n=1024$, $b = 4$, $r = 2^{10}$, $k = 0.45w$ and $t = 0.25$.}
  \centerline{\includegraphics[width=0.6\textwidth]{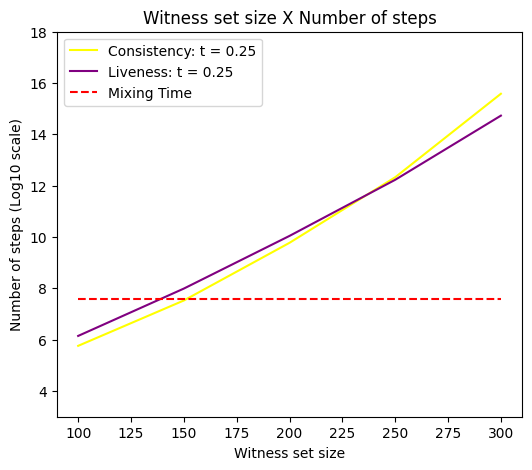}}
  \label{fig:liveness}
\end{figure}

To optimize the overall time of failure, one has to select the threshold $s$ so that it takes the same amount of steps to violate liveness and consistency.
In Figure~\ref{fig:liveness}, we show both curves for $s = 0.45$ and $t = 0.25$.\footnote{Ideally, $s$ should be selected to optimize the failure time for each value of $w$ and $t$. However, calculating liveness results takes considerably longer (than consistency) due to the size of matrix $P$. Therefore, we present the values for $t=0.25$, where the size of $P$ is smaller.}

\section{Recovery Protocol}
\label{ap:recovery}
\myparagraph{Protocol Description.} A process $p_i$ first sends $\RECOVER$ to every process and includes the latest $\WBB$ message it has sent with tag $\rPi$.
If a process $p_j$ that has already delivered a message $m$ receives $\RECOVER$, it replies with $m$.
When $p_i$ receives $f+1$ replies for $m$, it knows at least one is from a correct process, and can safely deliver it.
Moreover, when receiving $f+1$ $\RECOVER$ messages, $p_j$ also sends a $\RECOVER$ to every process (this threshold ensures that at least one correct process should propose to initiate a reliable broadcast instance).

When $p_i$ receives $\RECOVER$ from $\lfloor \frac{n+f}{2} \rfloor + 1$ processes, if a unique message $m \neq \perp$ was received so far, it then starts Bracha's broadcast $\ECHO$ phase for $m$.
Otherwise, $p_i$ waits until it receives $f+1$ $\langle \READY, m, \rPi \rangle$ from the recovery messages to start echoing.
The traditional Byzantine reliable broadcast is then executed.

\begin{algorithm}
\caption{Recovery Protocol}
\label{alg:recovery}
\BlankLine
\setcounter{AlgoLine}{17}

\nl \Guard{triggering $\Timeout$}{
    \nl \If{no message has been delivered}{
        \nl $m_{\Pi} \gets $ last message sent tagged with $\Pi$\;
        \Comment{$m_{\Pi} \gets \perp$ if none}
        \nl send $\langle \RECOVER, m_{\Pi} \rangle$ to every $p \in \Pi$\;
    }
} 
\BlankLine

\nl \Guard{receiving $\langle \RECOVER, m_{\Pi} \rangle$ from $p_j$}{
    \nl $\RecHist \gets \RecHist \cup \{m_{\Pi}\}$\;
    \nl \If{already delivered $m$}{
        \nl send $\langle \REPLY, m \rangle$ to $p_j$\;
    }
}
\BlankLine

\nl \Guard{receiving $f+1$ $\langle \REPLY, m \rangle$ messages}{ 
    \nl \If{no message has been delivered}{
        \nl $\deliver(m)$\; \label{line:recFirstDeliver}
    }
}
\BlankLine

\nl \Guard{receiving $\langle \RECOVER, \cdot \rangle$ from $f+1$ processes such that no $\RECOVER$ was sent}{
    \nl $m_{\Pi} \gets $ last message sent with tag $\rPi$\;
    \nl send $\langle \RECOVER, m_{\Pi} \rangle$ to every $p \in \Pi$\;
}
\BlankLine

\nl \Guard{receiving $\langle \RECOVER, \cdot \rangle$ from $\lfloor \frac{n+f}{2} \rfloor + 1$ processes such that no $\ECHO$ was sent}{
    \nl \label{line:recGuard1} \If{there is a unique $m \neq \perp$ such that $\langle \cdot, m, \rPi \rangle \in \RecHist$}{
        \nl send $\langle \ECHO, m \rangle$ to every $p \in \Pi$\; \label{line:recFirstEcho}
    }
}
\BlankLine

\nl \Guard{having $f+1$ $\langle \READY, m, \Pi \rangle \in \RecHist$ such that no $\ECHO$ was sent}{
    \nl send $\langle \ECHO, m \rangle$ to every $p \in \Pi$\; \label{line:recSecondEcho}
}

\nl \Guard{receiving $\lfloor \frac{n+f}{2} \rfloor +1$ $\langle \ECHO, m \rangle$ or $f+1$ $\langle \READY, m \rangle$ messages}{
    \nl send $\langle \READY, m \rangle$ to every $p \in \Pi$\;
}

\nl \Guard{receiving $\langle \READY, m \rangle$ from $\lfloor \frac{n+f}{2} \rfloor +1$ processes} {
    \nl \If{no message has been delivered}{
        \nl $\deliver(m)$\; \label{line:recSecondDeliver}
    }
}
\BlankLine
\end{algorithm}

\myparagraph{Correctness.} For a particular instance $f < |\Pi|/3$, we assume that for every correct process $p_i$, $W_i$ has at most $k-1$ faulty witnesses\footnote{Note that these assumptions are weaker than those of~\ref{subsec:correctness}. This is because the addition of the recovery protocol and a timeout compensate for the non-responsiveness of witness sets.}.
In addition, we assume the $\Timeout$ time to be set much smaller than the execution time of $\delta$ broadcast instances, such that the adversary cannot change the composition of faulty processes in witness sets before processes timeout.

\setcounter{theorem}{8}
\begin{lemma}
\label{lm:totalConsistency}
If two correct processes $p$ and $q$ deliver $m$ and $m'$, then $m = m'$.
\begin{proof}
There are three different scenarios according to the algorithm in which $p$ and $q$ deliver messages: both deliver in line~\ref{line:wbbDeliver} (Algorithm\ref{alg:wbb}), one delivers in line~\ref{line:wbbDeliver} and the other in lines~\ref{line:recFirstDeliver} or~\ref{line:recSecondDeliver} (Algorithm~\ref{alg:recovery}), or both deliver in lines~\ref{line:recFirstDeliver} or~\ref{line:recSecondDeliver}.

In the first case, since at most $k-1$ witnesses are faulty in each witness set, correct processes receive messages from at least one correct witness in lines~\ref{line:phase1k} and \ref{line:phase2k}, which guarantees $\Consistency$ (see proof of Theorem~\ref{th:sketchCorrectness}).

For the second case,
if $q$ delivers $m'$ in line~\ref{line:recFirstDeliver}, it is guaranteed to receive a reply from at least a correct process $r$.
Since processes only take step in Algorithm~\ref{alg:recovery} after timing out or delivering a message, it must be that $r$ delivered $m'$ in line~\ref{line:wbbDeliver}.
From the first case, $m = m'$.

On the other hand, if $q$ delivers $m'$ in line~\ref{line:recSecondDeliver}, it received $\lfloor \frac{n+f}{2} \rfloor + 1$ $\langle \READY, m' \rangle$.
Suppose that $m \neq m'$, then $\lfloor \frac{n+f}{2} \rfloor + 1$ processes sent $\langle \ECHO, m' \rangle$ (sufficient to make a correct process send ready for $m'$).
But since $p$ delivers $m$ after receiving $\langle \VALIDATE, m \rangle$ from at least a correct witness, it is also the case that at least $\lfloor \frac{n+f}{2} \rfloor + 1$ processes sent $\langle \READY, m, \rPi \rangle$.

Consequently, a correct process $r$ must have sent both $\langle \ECHO, m' \rangle$ and $\langle \READY, m, \Pi \rangle$.
Two scenarios are possible: if $r$ sent $\ECHO$ in line~\ref{line:recFirstEcho}, then it had readies for $m$ and $m'$ stored (since $r$ receives $\lfloor \frac{n+f}{2} \rfloor + 1$ recovery messages, as least one contains a $\langle \READY, m, \Pi \rangle$), a contradiction with the guard of line~\ref{line:recGuard1}.
If it was in line~\ref{line:recSecondEcho}, then a correct process sent $\langle \READY, m', \Pi \rangle$, also a contradiction since two correct processes sent $\langle \READY, \cdot, \Pi \rangle$ for distinct messages $m$ and $m'$ (see proof of Theorem~\ref{th:sketchCorrectness}).

In the third case, suppose $p$ delivers $m$ in line~\ref{line:recFirstDeliver} and $q$ delivers $m'$ in line\ref{line:recSecondDeliver}.
There is a correct process $r$ that delivers $m$ in line~\ref{line:wbbDeliver} and sent a reply to $p$, which from the second case above implies $m = m'$.
If both deliver $m$ and $m'$ in line~\ref{line:recSecondDeliver}, there is at least one correct process that send both $\langle \READY,m \rangle$ and $\langle \READY,m' \rangle$.
Since correct processes do not send readies for distinct messages, $m=m'$.
\end{proof}
\end{lemma}

\begin{lemma}
\label{lm:totalTotality}
If a correct process delivers a message, then every correct process eventually delivers a message.
\begin{proof} 
A correct process $p$ can deliver a message in three possible occasions: in line~\ref{line:wbbDeliver} (Algorithm~\ref{alg:wbb}), and lines~\ref{line:recFirstDeliver} and~\ref{line:recSecondDeliver} (Algorithm~\ref{alg:recovery}).

If $p$ delivers $m$ in line~\ref{line:wbbDeliver}, because $p$'s witness set has at least one correct witness that sends $\langle \ECHO, m, W \rangle$ to everyone, every correct process either times-out or delivers a message in line~\ref{line:wbbDeliver} (using witnesses).
Moreover, from Theorem~\ref{lm:totalConsistency}, no correct process delivers $m' \neq m$.
At least one correct witness $w$ sent $\langle \VALIDATE, m \rangle$ to $p$, $\lfloor \frac{n+f}{2} \rfloor + 1$ processes sent $\langle \READY, m, \Pi \rangle$ to $w$, from which at least $f+1$ are correct.

If another process $q$ times-out, it can deliver $m$ by receiving $f+1$ replies from a $\RECOVER$ message.
Suppose $q$ does not receive enough replies, then at least $f+1$ correct processes \emph{do not} deliver $m$ in line~\ref{line:wbbDeliver} (Algorithm~\ref{alg:wbb}), that is, they timeout.
Consequently, every correct process receives $f+1$ $\RECOVER$ messages and also send $\RECOVER$ (even if they already delivered $m$).
$q$ then gathers $\lfloor \frac{n+f}{2} \rfloor + 1$ $\RECOVER$ messages, and since there is at least one correct process among them that sent $\langle \READY, m, \Pi \rangle$, $q$ echoes $m$ (if $m$ is the only gathered message, line~\ref{line:recGuard1}).

If $q$ receives a distinct $m'$ before echoing a message, it waits for $f+1$ $\RECOVER$ messages containing $\langle \READY, m, \Pi \rangle$, which is guaranteed to happen (since at least $f+1$ correct processes previously sent $\langle \READY, m, \Pi \rangle$).
Thus, every correct process echoes $m$, gathers $\lfloor \frac{n+f}{2} \rfloor + 1$ echoes and sends $\langle \READY, m \rangle$.
Any process that has not delivered a message then receives $\lfloor \frac{n+f}{2} \rfloor + 1$ readies and deliver $m$.

In the case where $p$ delivers $m$ in line~\ref{line:recFirstDeliver}, at least one correct process sent reply to $p$ and delivered $m$ in Algorithm~\ref{alg:wbb}, which implies the situation described above.

Now suppose that $p$ delivers $m$ in line~\ref{line:recSecondDeliver}, and no correct process delivers a message in Algorithm~\ref{alg:wbb}.
$p$ received $\lfloor \frac{n+f}{2} \rfloor + 1$ $\langle \READY,m \rangle$, which at least $f+1$ are from correct processes.
Moreover, because correct processes wait for $\lfloor \frac{n+f}{2} \rfloor + 1$ echoes (or $f+1$ readies) before sending $\READY$, they cannot send $\READY$ for distinct messages $m$ and $m'$, since that would imply that a correct process sent $\ECHO$ for both messages.
Therefore, every correct process $q$ is able to receive $f+1$ readies for $m$ and also send ready for it.
$q$ then receives $\lfloor \frac{n+f}{2} \rfloor + 1$ readies and deliver $m$.
\end{proof}
\end{lemma}

\begin{lemma}
\label{lm:totalValidity}
If a correct process broadcasts $m$, every correct process eventually delivers $m$.
\begin{proof}
If any correct process delivers $m$, from Lemma~\ref{lm:totalTotality} every correct process delivers it.
Suppose that $p$ is the source and that no process delivers $m$ before it times-out.
$p$ then sends $\RECOVER$ (including $m$) to every process, so that even if $p$ reaches no correct witnesses, correct processes still receive a protocol message and initializes the instance.
Since no correct process delivers $m$ before timing-out, they also trigger $\Timeout$ and send $\RECOVER$ with $m$.

Because $p$ is correct, it sends no protocol message for $m'\neq m$. 
Every correct process then gathers enough $\RECOVER$ messages to send $\ECHO$ and later $\READY$.
$p$ eventually gathers $\lfloor \frac{n+f}{2} \rfloor + 1$ readies and deliver $m$.
\end{proof}
\end{lemma}

Lemmas~\ref{lm:totalConsistency},~\ref{lm:totalTotality} and~\ref{lm:totalValidity} imply Theorem~\ref{th:intPerformance}.

\begin{theorem}
\label{th:intPerformance}
Algorithms~\ref{alg:wbb},~\ref{alg:updates} and~\ref{alg:recovery} together satisfy $\Validity$, $\Consistency$ and $\Totality$.
\end{theorem}

\section{Applications and Ramifications}
\label{sec:applications}

In this section, we briefly overview two potential applications of our broadcast protocol: asynchronous asset transfer and a generic accountability mechanism. 
We also discuss open questions inspired by these applications.  

\subsection{Asset Transfer}

The users of an \emph{asset-transfer system} (or a \emph{cryptocurrency} system) exchange assets via \emph{transactions}. 
A transaction is a tuple $\tx = (s,r,v,\seq)$, where $s$ and $r$ are the sender's and receiver's \emph{account $\Id$} respectively, $v$ is the transferred amount and $\seq$ is a sequence number.
Each user $p$ maintains a local set of transactions $T$ and it adds a transaction $\tx$ to $T$ (we also say that $p$ \emph{commits} $\tx$), when $p$ confirms that (i)~all previous transactions from $s$ are committed, (ii)~$s$ has not issued a \emph{conflicting} transaction with the same sequence number, and (iii)~based on the currently committed transactions, $s$ indeed has the assets it is about to transfer.\footnote{Please refer to~\cite{astro-dsn} for more details.}

One can build such an abstraction atop (probabilistic) reliable broadcast: to issue a transaction $tx$, a user invokes $\broadcast(tx)$.
When a user receives an upcall $\deliver(tx)$ it puts $tx$ on hold until conditions (i)-(iii) above are met and then commits $tx$.
 Asset-transfer systems based on classical broadcast algorithms~\cite{bra87asynchronous} exhibit significant practical advantages over the consensus-based protocols~\cite{astro-dsn,fastpay}.
One can reduce the costs even further by using our broadcast protocol.
The downside is that there is a small probability of \emph{double spending}.
A malicious user may make different users deliver conflicting transactions and overspend its account by exploiting ``weak'' (not having enough correct members) witness sets or over-optimistic evaluation of communication delays in the recovery protocol. 
We can temporarily tolerate such an overspending and compensate it with a reconfiguration mechanism that detects and evicts misbehaving users from the system, as well as adjusting the total balance.  
It is appealing to explore whether such a solution would be acceptable in practice. 

\subsection{Accountability and beyond}

In our approach probabilistic protocol, every broadcast event is validated by a set of witnesses.
The validation here consists in ensuring that the source does not attempt to broadcast different messages with the same sequence number.
As a result, with high probability, all correct processes observe the same sequence of messages issued by a given source.  

One can generalize this solution to implement a lightweight accountability mechanism (in the vein of ~\cite{peerreview,fault-detection09}): witnesses collectively make sure that the sequence of events generated by a process corresponds to its specification.
Here a process commits not only to the messages it sends, but also to the messages it receives.
This way the witnesses may verify if its behavior respects the protocol the process is assigned.  

Notice that one can generalize this approach even further, as the verified events do not have to be assigned to any specific process.   
In the extreme case, we can even think of a probabilistic state-machine replication protocol~\cite{paxos,pbft}.
What if the processes try to agree on an ever-growing sequence of events generated by all of them in a decentralized way? 
Every next event (say, at position $k$) may then be associated with a dynamically determined pseudo-random set of witnesses that try to make sure that no different event is accepted at position $k$. 
Of course, we need to make sure that the probability of losing consistency and/or progress is acceptable and a probability analysis of this kind of algorithms is an appealing question for future research. 

\end{document}